\documentclass[11pt,a4paper]{article}
\pdfoutput=1

\usepackage[utf8]{inputenc}
\usepackage[T1]{fontenc}
\usepackage[margin=2.5cm]{geometry}
\usepackage[pdfstartview={FitH},linktoc=page,colorlinks=true,linkcolor=blue,citecolor=blue,urlcolor=blue]{hyperref}
\usepackage[numbered]{bookmark}
\usepackage{mathtools}
\usepackage{amssymb}
\usepackage{graphicx}
\usepackage[font=small,labelfont=bf]{caption}
\usepackage{authblk}
\usepackage{cite}
\usepackage{dsfont}
\usepackage{xcolor}
\usepackage[titles]{tocloft}

\numberwithin{equation}{section}
\allowdisplaybreaks

\newcommand*{\dd}{\mathop{}\!\mathrm{d}}

\begin{document}

\title{\vspace{2cm}\textbf{Non-abelian T-duality of Pilch-Warner background}\vspace{1cm}}
\author[*]{Hristo Dimov}
\author[*]{Stefan Mladenov}
\author[*,$\dagger$]{Radoslav C. Rashkov}
\author[*]{Tsvetan Vetsov}
\affil[*]{\textit{Department of Physics, Sofia University,}\authorcr\textit{5 J. Bourchier Blvd., 1164 Sofia, Bulgaria}\vspace{5pt}}
\affil[$\dagger$]{\textit{Institute for Theoretical Physics, Vienna University of Technology,}\authorcr\textit{Wiedner Hauptstr. 8--10, 1040 Vienna, Austria}\authorcr\vspace{10pt}\texttt{h\_dimov,smladenov,rash,vetsov@phys.uni-sofia.bg}\vspace{1cm}}
\date{}

\maketitle

\begin{abstract}
In this work we obtain the non-abelian T-dual geometry of the well-known Pilch-Warner supergravity solution in its infrared point. We derive the dual metric and the NS two-form by gauging the isometry group of the initial theory and integrating out the introduced auxiliary gauge fields. Then we use the Fourier-Mukai transform from algebraic geometry to find the transformation rules of the R-R fields. The dual background preserves the $\mathcal{N}=1$ supersymmetry of the original one due to the fact that the Killing spinor does not depend on the directions on which the N-AT-D is performed. Finally, we consider two different pp-wave limits of the T-dual geometry by performing Penrose limits for two light-like geodesics.
\end{abstract}

\vspace{1.5cm}

\thispagestyle{empty}
\newpage

\noindent\rule{\linewidth}{0.75pt}
\vspace{-0.8cm}\tableofcontents
\noindent\rule{\linewidth}{0.75pt}

\section{Introduction}

String theory occupied the attention of physics society in quest for a theory unifying all the fundamental interactions. The developments over years naturally led to the introduction of notion of D-branes as non-perturbative constituents of string theory. On other hand, the zoo of string dualities (S-, T-, U-, etc.), together with ultimate relations of branes with gauge theories and gravity, has opened many avenues for promising developments.

An example of widely used string duality is T-duality, which relates different string theories, for instance type IIA and type IIB, on spacetime backgrounds with reciprocal compactification radii. The main point is that, while one geometry has large radius $R$, the dual one has small radius $\alpha'/R$. The existence of such duality essentially means that there are two different mathematical descriptions of the same physical system. T-duality is also a perturbative duality in a sense that it relates the weak coupling regimes of both theories. This fact allows one to test it in perturbation theory via comparison of the corresponding string spectra.

In the abelian case there are different approaches to T-duality mainly due to the work of Buscher \cite{Buscher:1987sk,Buscher:1987qj}. In general, for any two-dimensional nonlinear sigma model with certain isometry group (abelian or not) there exists a clear procedure that produces the T-duality transformation rules. Firstly, one has to gauge the group structure by inserting Lagrange multipliers and auxiliary gauge fields into the Lagrangian of the theory to be dualised. The equations of motion for the Lagrange multipliers force the field strength to vanish. Secondly, substituted in the gauged Lagrangian, the solutions of those equations reproduce the original theory. Finally, one can integrate out the introduced auxiliary gauge fields (i.e. using path integral) and interpret the Lagrange multipliers as dual coordinates thus arriving at the Lagrangian of (non-)abelian T-dual geometry. Following this strategy in the abelian case, Ro\v{c}ek and Verlinde \cite{Rocek:1991ps} proved that T-duality is a symmetry between conformal field theories and can be realised as a quotient by combinations of chiral currents. They also extended their result to the case of $\mathcal{N}=2$ supersymmetric sigma models.

Unlike the abelian case, non-abelian T-duality (N-AT-D) \cite{delaOssa:1992vc} (for review see \cite{Giveon:1994fu}) is quite far from being completely understood. At the same time, its significance continues to kindle ongoing interest in physicists. This is mainly due to the fact that many models in theoretical physics possess non-abelian isometries. Additionally, N-AT-D can play a crucial role in the classification of inequivalent vacua in string theory. Although the procedure for deriving transformation rules of the metric and the antisymmetric Kalb-Ramond field is equivalent to the abelian case, N-AT-D has very distinctive properties. One remarkable feature is that starting with a geometry which has some isometry and apply N-AT-D on it, one typically obtains dual geometry in which this isometry is destroyed. Therefore it is impossible to apply this transformation procedure backwards---N-AT-D has an inverse transformation but, unless the isometry is $U(1)$, it is not its own inverse---despite the fact that the two geometries are connected by T-duality and their partition functions are equal. This issue is even more intriguing owing to the existence of duality symmetries for Calabi-Yau compactifications which do not have any isometries. Hence T-duality should be understood independently of any group structures of the underlying geometries.

Non-abelian T-duality poses some open questions which have to be thoroughly studied, and the answers of which may lead to new insights. In order to maintain conformal invariance, along with the transformations of the metric and the $B$-field one should transform appropriately the dilaton as well. However, it has been shown by a counter-example that this cannot be accomplished for a specific Bianchi-type cosmological model \cite{Gasperini:1993nz}. More surprisingly, it does not exist any dilaton transformation which is able to satisfy the $\beta$-function equations and restore the conformal invariance in the dual theory. In contrast to the abelian case, where the Lagrange multiplier is singlet under the action of the gauge group and can be naturally made periodic, for non-abelian gauge groups the Lagrange multipliers transform in the adjoint representation and the group action prevent one introducing winding modes \cite{Giveon:1993ai}. For theories with abelian isometries the action is invariant under constant shifts of the Lagrange multiplier, which ensures that the dual theory also has symmetry to be gauged and, repeating the T-duality procedure, one can restore the original model. However, this is no longer the case for N-AT-D; the action symmetry might become nonlocal, hence performing second T-duality could lead to completely different theory than the original one. Thus one can infer that N-AT-D is not a symmetry of a conformal field theory, but a symmetry between different conformal field theories.

Together with the metric and the NS two-form\footnote{The massless string spectrum contains also dilaton field, which vanishes for the Pilch-Warner geometry of interest in the present paper.}, the R-R fields also transform under abelian T-duality. Various methods have been developed for performing this transformation. Amongst them are (without pretending to be exhaustive): demonstrating that the actions of type II supergravities are equivalent after reduction to nine dimensions \cite{Bergshoeff:1995as,Meessen:1998qm}; using R-R vertex operators \cite{Polchinski:1996na}; obtaining T-duality transformations of spacetime spinors and requiring consistency between T-duality and supersymmetry \cite{Hassan:1999bv}; employing pure spinor formalism and worldsheet path integral derivation \cite{Benichou:2008it}. The transformation of the R-R sector under non-abelian T-duality closely mirrors the abelian cases. The first paper, which considered N-AT-D in R-R backgrounds in the context of holography and triggered the recent developments in the subject, is \cite{Sfetsos:2010uq}. The corresponding non-abelian versions of the above methods are: reductions to seven dimensions on round and fibred round three-spheres \cite{Itsios:2012dc,Jeong:2013jfc}; generalisation of Hassan's method \cite{Itsios:2013wd,Kelekci:2014ima}; transformation of six-dimensional internal spinor constructed from $G$-structure \cite{Barranco:2013fza,Macpherson:2015tka}; generalisation to the cases in which the isometry is a coset \cite{Lozano:2011kb}.

In this paper we adopt a different approach. T-duality can be lifted to K-theory as a Fourier-Mukai transform with Poincar\'e bundle as a kernel. It has been shown \cite{Hori:1999me} that Nahm transform of instantons is related to the transformation of D-branes under T-duality and the concrete isomorphism between K-theory groups, which realises this transformation on a torus of arbitrary dimensions, has been identified. On the other hand, the equations of motion for a CFT with topological defects imply a direct connection between the Poincar\'e line bundle and the boundary conditions affected by T-duality \cite{Sarkissian:2008dq}. Since boundary conditions correspond to D-branes (which are sources of R-R charges), one can define an action of a topological defect on the R-R charges. This action is of Fourier-Mukai type with kernel given by the gauge invariant flux of the defect \cite{Gevorgyan:2013xka}.

Our work is also motivated by the achievements of the AdS/CFT correspondence \cite{Maldacena:1997re}. This is a framework providing a successful non-perturbative approach to strongly coupled Yang-Mills theories via the properties of their dual classical supergravity solutions and vice versa. It is well-known fact that, at low energies, quantum chromodynamics---the most successful theory of strong interactions---is also strongly coupled gauge theory and does not allow any perturbative treatments. This forces one to look for an alternative non-perturbative description of the theory in the context of the gauge/gravity duality. Anyhow, in order to produce more realistic QCD-like string models, one has to find highly non-trivial background solutions that break significantly the amount of supersymmetry and conformal symmetry. An example of such non-trivial supergravity background is the Pilch-Warner geometry \cite{Pilch:2000ej}, which is holographically dual to $\mathcal{N}=1$ Leigh-Strassler theory \cite{Leigh:1995ep}.

This paper is organised as follows. In section \ref{sec:N-AT-D-gs} we give a brief review of the general setup of non-abelian T-duality following the method of gauging the non-abelian background isometries. Then in section \ref{sec:N-AT-D-PW} we consider the special case of Pilch-Warner supergravity solution. The background geometry consists of warped AdS spacetime times warped squashed sphere and, as we mentioned above, string theory on this background is of interest for AdS/CFT correspondence. We apply the N-AT-D to the subspace of the squashed sphere which has manifest $SU(2)$ symmetry. In section \ref{sec:TopDefAndF-MOfR-R} we introduce topological defects and using them perform Fourier-Mukai transform on the R-R fields. Section \ref{sec:susy} comprises analysis of the amount of supersymmetry preserved under N-AT-D based on the Kosmann spinorial Lie derivative. To cover some implications for the dual gauge theory, i.e. BMN operators, we present in section \ref{sec:pp-wave} two different pp-wave limits of the T-dual geometry. Finally, in section \ref{sec:conclusion} we conclude with a short summary of our results.

\section{Non-abelian T-duality---general setup}\label{sec:N-AT-D-gs}

We begin our consideration with a brief review of the general setup of non-abelian T-duality \cite{Gevorgyan:2013xka} (see also \cite{Itsios:2013wd,Sfetsos:2013wia,Bea:2015fja}). Initially, we need to explicitly write down the general transformation rules under N-AT-D for a given background with metric $\dd s^2$ and non-trivial NS two-form $B$. The group structure supported by the background is $G$, with generators $T^a$ and structure constants $f_{bc}^a$, $a=1,\ldots,\dim(G)$. The coordinates $\theta^k$ describe the $G$-part of the background. Let us use for the metric and the $B$ field the notations
\begin{align}
\dd s^2&=G_{\mu\nu}(Y)\dd Y^\mu\dd Y^\nu+2G_{\mu a}\Omega_k^a\dd Y^\mu\dd\theta^k+
G_{ab}\Omega_m^a\Omega_k^b\dd\theta^m\dd\theta^k,\\
B&=\frac{1}{2}B_{\mu\nu}(Y)\dd Y^\mu\dd Y^\nu+B_{\mu a}\Omega_k^a\dd Y^\mu\dd\theta^k+
\frac{1}{2}B_{ab}\Omega_m^a\Omega_k^b\dd\theta^m\dd\theta^k.
\end{align}
For any group element $g\in G$ one can construct the Maurer-Cartan form and further decompose it over the generators
\begin{equation}
g^{-1}\dd g=L^aT_a=\Omega_k^a\dd\theta^kT_a,
\end{equation}
where the left-invariant one-form fields $L^a$ satisfy the Maurer-Cartan equation\footnote{It follows from $\dd\left(g^{-1}\dd g\right)=-g^{-1}\dd g\wedge g^{-1}\dd g$.}
\begin{align}
&\dd L^a=-\frac{1}{2}f_{bc}^aL^bL^c,\quad\text{or}\label{eq:maurer-cartan}\\
&\partial_i\Omega_j^c-\partial_j\Omega_i^c=-f_{ab}^c\Omega_i^a\Omega_j^b.
\end{align}
In accordance with the Buscher procedure it is convenient to use the following form of the Lagrangian:
\begin{equation}
\label{eq:L1}
\mathcal{L}=Q_{\mu\nu}\,\partial Y^\mu\bar{\partial}Y^\nu+Q_{\mu a}\Omega_k^a\,\partial Y^\mu\bar{\partial}\theta^k+
Q_{a\mu}\Omega_k^a\,\partial\theta^k\bar{\partial}Y^\mu+Q_{ab}\Omega_m^a\Omega_k^b\,\partial\theta^m\bar{\partial}\theta^k,
\end{equation}
where
\begin{equation}
Q_{\mu\nu}=G_{\mu\nu}+B_{\mu\nu},\quad Q_{\mu a}=G_{\mu a}+B_{\mu a},\quad Q_{ab}=G_{ab}+B_{ab}.
\end{equation}

The next essential step is to gauge the group structure by introducing Lagrange multipliers\footnote{In this case the auxiliary gauge fields and the Lagrange multipliers take values in the Lie algebra associated to the group $G$.} $x^a$ and gauge fields $A^a$ in the Lagrangian of the original theory
\begin{align}
\label{eq:gaugedL1}
\mathcal{L}=Q_{\mu\nu}\,\partial Y^\mu\bar{\partial}Y^\nu+Q_{\mu a}\,\partial Y^\mu\bar{A}^a&+Q_{a\mu}A^a\,\bar{\partial}Y^\mu+Q_{ab}A^a\bar{A}^b\nonumber\\*
&-x^a(\partial\bar{A}^a-\bar{\partial}A^a+f_{bc}^aA^b\bar{A}^c).
\end{align}
Using the already gauged Lagrangian one can easily reproduce the original theory by simply eliminating the Lagrangian multipliers $x^a$. The equations of motion $\delta\mathcal{L}/\delta x^a=0$ imply vanishing field strength
\begin{equation}
F_{+-}^a=\partial\bar{A}^a-\bar{\partial}A^a+f_{bc}^aA^b\bar{A}^c=0,
\end{equation}
which has the obvious solutions
\begin{equation}
\label{eq:A1}
A^a=\Omega_k^a\,\partial\theta^k,\quad\bar{A}^a=\Omega_k^a\,\bar{\partial}\theta^k.
\end{equation}
Plug in back these solutions into \eqref{eq:gaugedL1} we get back to the original theory \eqref{eq:L1}.

On the other hand, to obtain the T-dual theory we have to integrate out the gauge fields $A^a$. This leads to the conditions (up to boundary terms)
\begin{align}
Q_{\mu a}\,\partial Y^\mu+Q_{ba}A^b-x^cf_{ba}^cA^b+\partial x^a&=0,\\
Q_{a\mu}\,\bar{\partial}Y^\mu+Q_{ab}\bar{A}^b-x^cf_{ab}^c\bar{A}^b-\bar{\partial}x^a&=0.
\end{align}
The solutions of these conditions are
\begin{align}
A^a&=-M_{ba}^{-1}(Q_{\mu b}\,\partial Y^\mu+\partial x^b),\label{eq:A2}\\
\bar{A}^a&=M_{ab}^{-1}(\bar{\partial}x^b-Q_{b\mu}\,\bar{\partial}Y^\mu)\label{eq:A3},
\end{align}
where we have defined
\begin{equation}
M_{ab}=Q_{ab}-x^cf_{ab}^c.
\end{equation}
To obtain the N-AT-D theory we plug in $A^a$ and $\bar{A}^a$ back into \eqref{eq:gaugedL1}. The resulting Lagrangian is\footnote{Many terms vanish because of the structure (antisymm)$\cdot$(symm).}
\begin{equation}\label{eq:L2}
\hat{\mathcal{L}}=\widehat{E}_{\mu\nu}\,\partial Y^\mu\bar{\partial}Y^\nu+\widehat{E}_{\mu a}\,\partial Y^\mu\bar{\partial}x^a+ \widehat{E}_{a\mu}\,\partial x^a\bar{\partial}Y^\mu+\widehat{E}_{ab}\,\partial x^a\bar{\partial}x^b,
\end{equation}
where
\begin{equation}
\widehat{E}_{\mu\nu}=Q_{\mu\nu}-Q_{\mu a}M_{ab}^{-1}Q_{b\nu},\quad\widehat{E}_{\mu a}=Q_{\mu b}M_{ba}^{-1},\quad \widehat{E}_{a\mu}=-Q_{b\mu}M_{ab}^{-1},\quad\widehat{E}_{ab}=M_{ab}^{-1}.
\end{equation}

The quantities of the two pictures are related by \eqref{eq:A1} and \eqref{eq:A2}, \eqref{eq:A3}. The comparison gives
\begin{align}
\Omega_k^a\,\partial\theta^k&=-M_{ba}^{-1}(Q_{\mu b}\,\partial Y^\mu+\partial x^b),\\
\Omega_k^a\,\bar{\partial}\theta^k&=M_{ab}^{-1}(\bar{\partial}x^b-Q_{b\mu}\,\bar{\partial}Y^\mu).
\end{align}
Finally, the dual field content in the NS sector can be extracted by separating symmetric and antisymmetric parts of the quantity $\widehat{E}_{AB}$. The result is:
\begin{align}
\widehat{G}_{\mu\nu}&=G_{\mu\nu}-\frac{1}{2}\left(Q_{\mu a}M_{ab}^{-1}Q_{b\nu}+Q_{\nu a}M_{ab}^{-1}Q_{b\mu}\right),\label{eq:dualGmunu}\\
\widehat{G}_{\mu a}&=\frac{1}{2}\left(Q_{\mu b}M_{ba}^{-1}-Q_{b\mu}M_{ab}^{-1}\right),\label{eq:dualGmua}\\
\widehat{G}_{ab}&=\frac{1}{2}\left(M_{ab}^{-1}+M_{ba}^{-1}\right),\label{eq:dualGab}\\
\widehat{B}_{\mu\nu}&=B_{\mu\nu}-\frac{1}{2}\left(Q_{\mu a}M_{ab}^{-1}Q_{b\nu}-Q_{\nu a}M_{ab}^{-1}Q_{b\mu}\right),\label{eq:dualBmunu}\\
\widehat{B}_{\mu a}&=\frac{1}{2}\left(Q_{\mu b}M_{ba}^{-1}+Q_{b\mu}M_{ab}^{-1}\right),\label{eq:dualBmua}\\
\widehat{B}_{ab}&=\frac{1}{2}\left(M_{ab}^{-1}-M_{ba}^{-1}\right).\label{eq:dualBab}
\end{align}
The above expressions define a N-AT-D procedure which gives the dual of geometry possessing some non-abelian group of symmetry $G$ acting without isotropy. The group structure is implicitly encoded in the matrix $M_{ab}$ through the structure constants $f_{ab}^c$. The fact that the group acts without isotropy is crucial because this allows us to completely fix the gauge by algebraic conditions on the target space coordinates. Consequently, we are able to clearly distinguish between dual and original coordinates. In the next section we will use this procedure to obtain the N-AT-D for the particular case of $SU(2)$ group structure supported by the Pilch-Warner geometry, namely we will dualise the celebrated Pilch-Warner supergravity solution.

\section{N-AT-D for Pilch-Warner}\label{sec:N-AT-D-PW}

The Pilch-Warner (PW) geometry \cite{Pilch:2000ej,Pilch:2000ue,Pilch:2000fu,Brecher:2002ar,Dimov:2003bh} is a solution to five-dimensional $\mathcal{N}=8$ supergravity lifted to ten dimensions. It is gravity dual to $\mathcal{N}=4$ gauge theory softly broken to $\mathcal{N}=2$ \cite{Buchel:2000cn,Buchel:2013id}. The solution preserves 1/8 of the initial supersymmetry over the flow except in the UV and IR fixed points. In the UV point the solution coincide with the maximally supersymmetric $\text{AdS}_5\times S^5$ solution, while in the IR fixed point it gives a geometry, which is a direct product of warped $\text{AdS}_5$ and squashed $S^5$, that preserves 1/4 of the initial supersymmetry and also has additional $SU(2)$ symmetry\footnote{To be more precise the metric has a global isometry group $SU(2)\times U(1)_\beta\times U(1)_\phi$.}. In this section we are especially interested in dualising the geometry in the IR critical point, hence all considerations hereafter correspond to this point of the flow.

The $SU(2)$ group structure, supported by the background, is parameterised by the left-invariant one-forms $\sigma_i,~i=1,2,3$ (for simplicity, 1/2 is added on to their standard definition):
\begin{align}\label{eq:left-inv-forms}
\sigma_1&=\frac{1}{2}(\sin\beta\dd\alpha-\cos\beta\sin\alpha\dd\gamma),\nonumber\\
\sigma_2&=-\frac{1}{2}(\cos\beta\dd\alpha+\sin\beta\sin\alpha\dd\gamma),\nonumber\\
\sigma_3&=\frac{1}{2}(\dd\beta+\cos\alpha\dd\gamma).
\end{align}
Using this definition and global coordinates we can write down the IR-point PW metric in the co-frame of left-invariant one-forms as follows:
\begin{align}
\dd s_{1,4}^2(\text{IR})&=L^2\Omega^2\left(-\cosh^2\rho\dd\tau^2+\dd\rho^2+\sinh^2\rho\dd\Omega_3^2\right),\\
\dd s_5^2(\text{IR})&=\frac{2}{3}L^2\Omega^2\left[\vphantom{\left(\dd\phi-\frac{4\cos^2\theta}{1-3\cos2\theta}\sigma_3\right)^2}\dd\theta^2+\frac{4\cos^2\theta}{3-\cos2\theta}\left(\sigma_1^2+\sigma_2^2\right) +\frac{4\sin^22\theta}{(3-\cos2\theta)^2}\left(\sigma_3+\dd\phi\right)^2\right.\nonumber\allowdisplaybreaks[0]\\
&\left.+\;\frac{2}{3}\left(\frac{1-3\cos2\theta}{\cos2\theta-3}\right)^2\left(\dd\phi-\frac{4\cos^2\theta}{1-3\cos2\theta}\sigma_3\right)^2\right],
\end{align}
where $\dd\Omega_3^2=\dd\phi_1^2+\sin^2\phi_1\left(\dd\phi_2^2+\sin^2\phi_2\dd\phi_3^2\right)$ is the metric of unit 3-sphere. The radius of the AdS throat in the IR point, $L$, is related to its counterpart in the UV point, $L_0$, by ${L=3\times2^{-5/3}L_0}$. The warp factor $\Omega^2$ is given by
\begin{equation}
\Omega^2=\frac{2^{1/3}}{\sqrt{3}}\sqrt{3-\cos2\theta}.
\end{equation}
As a supergravity solution the PW background contains non-trivial Kalb-Ramond two-form $B$:
\begin{equation}
B=-\frac{4}{9}2^{1/3}L^2\cos\theta\left[\dd\theta\wedge\sigma_1+\frac{2\sin2\theta}{3-\cos2\theta}(\dd\phi+\sigma_3) \wedge\sigma_2\right].
\end{equation}

The group element $g\in SU(2)$ can be parametrised in terms of left-invariant one-forms as follows:
\begin{equation}
g^{-1}\dd g=
\begin{pmatrix}
i\sigma_3 & \sigma_2+i\sigma_1 \\
-\sigma_2+i\sigma_1 & -i\sigma_3
\end{pmatrix},\quad
T_a=i\tau_a,\quad [T_a,T_b]=-2\epsilon_{ab}^cT_c,
\end{equation}
where $\epsilon_{ab}^c$ is the totally antisymmetric Levi-Civita symbol, the group generators $T_a$ are defined via the Pauli matrices $\tau_a$, and the structure constants are $f_{ab}^c=-2\epsilon_{ab}^c$. In the sigma co-frame $\dd\theta^k=\sigma_k$ the $\Omega_{3\times3}$ matrix simply becomes the identity matrix
\begin{equation}
\Omega_{3\times3}=\mathds{1}_{3\times3}.
\end{equation}
We can now compute the $M_{ab}$ matrix. For the sake of simplicity we define this matrix as in the following way
\begin{equation}
M_{ab}=Q_{ab}+2L^2\epsilon_{ab}^cx_c,
\end{equation}
where we included a factor of $L^2$ in the r.h.s. to ensure the correct dimensionality of the whole expression. The explicit form of the matrix $M_{ab}$ for the PW background is
\begin{equation}
M_{ab}=2L^2
\begin{pmatrix}
\frac{4\times2^{2/3}\cos^2\theta}{9\Omega^2} & x_3 & -x_2 \\
-x_3 & \frac{4\times2^{2/3}\cos^2\theta}{9\Omega^2} & x_1+\frac{16\cos^2\theta\sin\theta}{27\Omega^4} \\
x_2 & -x_1-\frac{16\cos^2\theta\sin\theta}{27\Omega^4} & \frac{16\times2^{1/3}\cos^2\theta(5-\cos2\theta)}{81\Omega^6}
\end{pmatrix}.
\end{equation}
In order to write down the final result in more compact form, it is convenient also to define the quantity
\begin{equation}
M=\frac{3^7\,\Omega^{10}}{2^5\,L^6\cos^2\theta}\det(M_{ab}),
\end{equation}
or explicitly
\begin{align}
M=108\times2^{1/3}\Omega^4x_3^2(5-\cos2\theta)+2^{2/3}\Omega^4\left[243\Omega^4(x_1^2+x_2^2)+ 256\times2^{-2/3}\cos^4\theta\right.\nonumber\\
\left.+288x_1\cos^2\theta\sin\theta\right].
\end{align}
At the end of the day, using formulas \eqref{eq:dualGmunu}--\eqref{eq:dualBab} we are in a position to compute all non-zero components of the non-abelian T-dual metric and Kalb-Ramond field. As expected, the warped AdS part of the original background remains unaffected under N-AT-D:
\begin{equation}
\widehat{\dd s_{1,4}^2(\text{IR})}=\dd s_{1,4}^2(\text{IR}).
\end{equation}
Nevertheless, the squashed five-sphere transforms under N-AT-D in a much more complex manner. Equation \eqref{eq:dualGmunu} produces the following non-zero components of the dual metric:
\begin{align}
&\widehat{G}_{\theta\theta}=G_{\theta\theta}+\frac{2\times2^{2/3}L^2\Omega^6}{9M}(243\Omega^4x_1^2+ 256\times2^{-2/3}\cos^4\theta+288\cos^2\theta\sin\theta),\nonumber\\
&\widehat{G}_{\theta\phi}=\widehat{G}_{\phi\theta}=\frac{16\times2^{1/3}L^2\Omega^2x_2\cos\theta}{3M}(10+8\cos2\theta- 2\cos4\theta+27\Omega^4x_1\sin\theta),\nonumber\\
&\widehat{G}_{\phi\phi}=G_{\phi\phi}-\frac{64L^2\cos^2\theta}{243M\Omega^6}\left[(256+729\Omega^8x_2^2)\cos2\theta+ 80\cos4\theta-4\cos8\theta\right.\nonumber\\
&\phantom{\hat{G}_{\phi\phi}=G_{\phi\phi}}\left.+9(20+72\times2^{2/3}\Omega^4x_3^2-81\Omega^8x_2^2+ 192\Omega^4x_1\cos^2\theta\sin\theta)\right].
\end{align}
Furthermore, the duality procedure \eqref{eq:dualGmua} mixes all Lagrange multipliers $x_a$, interpreted as new dual coordinates, with the $\theta$ and $\phi$ directions of the initial geometry:
\begin{align}
&\widehat{G}_{\theta x_1}=-\frac{L^2\Omega^2\cos^3\theta}{3\times2^{2/3}M}(768-256\cos2\theta+ 729\Omega^8x_1^2\sec^4\theta+864\Omega^4x_1\sec\theta\tan\theta),\nonumber\\
&\widehat{G}_{\theta x_2}=-\frac{L^2\Omega^4}{2^{2/3}M}\left[243\Omega^6x_1x_2\sec\theta+ 48\cos\theta\left(2^{1/3}x_3(5-\cos2\theta)+3\Omega^2x_2\sin\theta\right)\right],\nonumber\\
&\widehat{G}_{\theta x_3}=-\frac{9L^2\Omega^6\cos\theta}{2^{2/3}M}(-12\times2^{2/3}\Omega^2x_2+ 27\Omega^4x_1x_3\sec^2\theta+16x_3\sin\theta),\nonumber\\
&\widehat{G}_{\phi x_1}=-\frac{2L^2\cos\theta\sin2\theta}{M}\left[-16\times2^{1/3}x_3(5-\cos2\theta)+ 81\Omega^6x_1x_2\sec^2\theta+48\Omega^2x_2\sin\theta\right],\nonumber\\
&\widehat{G}_{\phi x_2}=-\frac{2L^2\sec\theta\sin2\theta}{9\Omega^2M}\left[729\Omega^8x_2^2+ 128\cos^4\theta(5-\cos2\theta)\right],\nonumber\\
&\widehat{G}_{\phi x_3}=-\frac{2L^2\sec\theta\sin2\theta}{3M}\left[243\Omega^6x_2x_3+ 4\times2^{2/3}\cos^2\theta(27\Omega^4x_1+16\cos^2\theta\sin\theta)\right].
\end{align}
The remaining part of the dual geometry is generated by eq. \eqref{eq:dualGab} and is comprised of the following metric components:
\begin{align}
&\widehat{G}_{x_1x_1}=\frac{3L^2\Omega^2\cos^2\theta}{8M}(768-256\cos2\theta+729\Omega^8x_1^2\sec^4\theta+ 864\Omega^4x_1\sec\theta\tan\theta),\nonumber\\
&\widehat{G}_{x_2x_2}=\frac{3L^2\Omega^2\sec^2\theta}{8M}\left[729\Omega^8x_2^2+128\cos^4\theta(5- \cos2\theta)\right],\nonumber\\
&\widehat{G}_{x_3x_3}=\frac{27L^2\Omega^6\sec^2\theta}{8M}(81\Omega^4x_3^2+32\times2^{1/3}\cos^4\theta),\nonumber\\
&\widehat{G}_{x_1x_2}=\widehat{G}_{x_2x_1}=\frac{81L^2\Omega^6x_2}{8M}(27\Omega^4x_1\sec^2\theta+16\sin\theta),\nonumber\\
&\widehat{G}_{x_1x_3}=\widehat{G}_{x_3x_1}=\frac{81L^2\Omega^6x_3}{8M}(27\Omega^4x_1\sec^2\theta+16\sin\theta),\nonumber\\
&\widehat{G}_{x_2x_3}=\widehat{G}_{x_3x_2}=\frac{2187L^2\Omega^{10}x_2x_3\sec^2\theta}{8M}.
\end{align}
The dual NS two-form has significantly simpler form. Formula \eqref{eq:dualBmunu} contributes with only one non-zero component
\begin{equation}
\widehat{B}_{\theta\phi}=-\widehat{B}_{\phi\theta}=\frac{32\times2^{2/3}L^2x_3\cos\theta}{9M}(27\Omega^4x_1-6\sin\theta- 5\sin3\theta+\sin5\theta),
\end{equation}
while \eqref{eq:dualBmua} does not generate $\theta x_a$-mixed terms:
\begin{align}
&\widehat{B}_{\phi x_1}=\frac{8\times2^{1/3}L^2}{M}\left[27\Omega^4x_1x_3+4\cos^2\theta(3\times2^{2/3}\Omega^2x_2+ 4x_3\sin\theta)\right],\nonumber\\
&\widehat{B}_{\phi x_2}=\frac{8L^2}{9\Omega^2M}\left[243\times2^{1/3}\Omega^6x_2x_3-8\cos^2\theta(27\Omega^4x_1+ 4\sin\theta+4\sin3\theta)\right],\nonumber\\
&\widehat{B}_{\phi x_3}=\frac{8\times2^{1/3}L^2}{3M}(81\Omega^4x_3^2+32\times2^{1/3}\cos^4\theta),
\end{align}
and eq. \eqref{eq:dualBab} generates all possible $x_ax_b$-mixed terms:
\begin{align}
&\widehat{B}_{x_2x_1}=-\widehat{B}_{x_1x_2}=\frac{54\times2^{1/3}L^2\Omega^4x_3}{M}(5-\cos2\theta),\nonumber\\
&\widehat{B}_{x_3x_2}=-\widehat{B}_{x_2x_3}=\frac{9L^2\Omega^4}{2^{1/3}M}(27\Omega^4x_1+4\sin\theta+4\sin3\theta),\nonumber\\
&\widehat{B}_{x_1x_3}=-\widehat{B}_{x_3x_1}=\frac{243L^2\Omega^8x_2}{2^{1/3}M}.
\end{align}
It is worthy to mention that all dual quantities have proper scaling, i.e. all of them are proportional to $L^2$. This is a consequence of the insertion of $L^2$ factor in the definition of the quantity $M$. In the next section we proceed with the transformation rules of the R-R fields under N-AT-D.

\section{Fourier-Mukai transform of the R-R fields via topological defects}\label{sec:TopDefAndF-MOfR-R}

In the context of two-dimensional quantum field theory defects can be considered as oriented lines separating two different quantum field theories. Using the left- and right-moving energy-momentum tensors of the two theories (denoted here by 1 and 2) $T^{(1)}$, $T^{(2)}$ and $\bar{T}^{(1)}$, $\bar{T}^{(2)}$ respectively, one can basically distinguish between two types of defects. Conformal defects satisfy the condition $T^{(1)}-T^{(2)}=\bar{T}^{(1)}-\bar{T}^{(2)}$ whereas topological defects are defined by the equalities $T^{(1)}=\bar{T}^{(1)}$ and $T^{(2)}=\bar{T}^{(2)}$. Remembering that the energy-momentum tensor generates diffeomorphisms, it is apparent that topological defects are invariant under deformations of the line they are affixed to. In this section we are particularly interested in those defects because they have the feature to be moved without changing the correlator \cite{Petkova:2001zn}. Thus one can define the notion of fusion between a topological defect and a boundary.

In what follows, we will review the implementation of topological defects in sigma model actions \cite{Sarkissian:2008dq}. Let us place a topological defect $Z$ on the line $\sigma=0$ of a two-dimensional oriented manifold (worldsheet) $\Sigma$. As a result, we have two connected regions $\Sigma_1$ ($\sigma\leq 0$) and $\Sigma_2$ ($\sigma\geq 0$). The pair of maps $X:\Sigma_1\rightarrow M_1$ and $Y:\Sigma_2\rightarrow M_2$ provides sigma model description for two theories with target spaces $M_1$ and $M_2$ respectively. The correct worldvolume description of topological defects is in terms of objects called bi-branes \cite{Fuchs:2007fw}; these are submanifolds of the Cartesian product of the two target spaces: $Q\subset M_1\times M_2$. The combined map of $(X,Y)$ on the defect should take values in $Q$:
\begin{align}
\Phi:~Z&\rightarrow M_1\times M_2\nonumber\\
z&\mapsto(X(z),Y(z)).
\end{align}
The submanifold $Q$ is endowed with a vector bundle (in the case of non-trivial $B$-field this is a twisted vector bundle or, equivalently, a gerbe bimodule) and a connection one-form $A$. The worldsheet action in presence of topological defects can be written down as
\begin{equation}\label{eq:defaction}
S=\int_{\Sigma_1}L_1+\int_{\Sigma_2}L_2+\int_Z\Phi^*A,
\end{equation}
where $\Phi^*$ is the pullback of $\Phi$ and $L_i=E^{(i)}_{mn}\partial X^m\bar{\partial}X^n$, $E^{(i)}=G^{(i)}+B^{(i)}$, $i=1,2$, are the usual bulk Lagrangians of the two theories. Naively thinking, the target space of the sigma model written this way looks thirteen-dimensional, however this is not the case. In fact, the sigma model is still ten-dimensional and has as target space coordinates the original $Y^\mu$ and $\theta^k$. The three new $x^a$'s are still interpreted as Lagrange multipliers. They become new coordinates not before after the $SU(2)$ group coordinates $\theta^k$ are integrated out in order to obtain the non-abelian T-dual theory. Thus, the sigma model is always kept ten-dimensional and conformally invariant.

The action with defects \eqref{eq:defaction} allows straightforward implementation of non-abelian T-duality. The setup is as follows. The target space $M_1$ is identified with the target space of the original geometry with coordinates $(Y^\mu,\theta^k)$ and the target space $M_2$ is identified with the N-AT-D geometry with coordinates $(Y^\mu,x^a)$. This means that the Lagrangians $L_1$ and $L_2$ in \eqref{eq:defaction} are given by \eqref{eq:L1} and \eqref{eq:L2} respectively. The bi-brane $Q$ has coordinates $(Y^\mu,\theta^k,x^a)$ and connection
\begin{equation}
A=-x^aL^a=-x^a\Omega_k^a\dd\theta^k,
\end{equation}
where, as before, $L^a$ are the left-invariant Maurer-Cartan one-forms. Consequently, we can define a non-abelian Poincar\'e line bundle $\mathcal{P}^\text{NA}$ over $Q$ with curvature
\begin{equation}\label{eq:curvature}
F=\dd A=-\dd x^aL^a+\frac{1}{2}x^af_{bc}^aL^bL^c.
\end{equation}
The second term in the above equation has been obtained using the Maurer-Cartan equation \eqref{eq:maurer-cartan}. One can easily check that the solutions of the equations of motion for the defect line in \eqref{eq:defaction} reproduce the transformation rules of N-AT-D derived in section \ref{sec:N-AT-D-gs} (the specific calculations can be found in section 3.2 of \cite{Gevorgyan:2013xka}). Moreover, they give the necessary conditions for the defect to be topological.

Armed with this set-up, we are ready to proceed with the transformation of R-R field strengths under non-abelian T-duality. As previously mentioned, topological defects can be moved and fused with a boundary, thus causing a change in the boundary conditions. The boundary conditions on their own give rise to D-branes, which are sources of R-R charges or elements of K-theory. Hence topological defects generate transformations of the R-R field strengths. These transformations (for the case of toroidal compactifications) are suggested to be given by Fourier-Mukai transform \cite{Hori:1999me,Sarkissian:2008dq} with kernel $e^\mathcal{F}$, where $\mathcal{F}=\widehat{B}-B+F$ is the gauge invariant two-form flux on the defect. For the purposes of N-AT-D, we define the Fourier-Mukai transform in the following way. Suppose that for two target spaces $M_1$ and $M_2$ we can define two corresponding rings $R(M_1)$ and $R(M_2)$ in such a way that for a map $p:M_1\rightarrow M_2$ there exist pullback $p^*:R(M_2)\rightarrow R(M_1)$ and pushforward $p_*:R(M_1)\rightarrow R(M_2)$ maps. We are also equipped with two projections $p^{M_1}:M_1\times M_2\rightarrow M_1$ and $p^{M_2}:M_1\times M_2\rightarrow M_2$, and an element $K\in R(M_1\times M_2)$. Then, for $F\in R(M_1)$, one can define the Fourier-Mukai transform $\text{FM}(F):R(M_1)\rightarrow R(M_2)$ with kernel $K$ in the following way:
\begin{equation}
\text{FM}(F)=p_*^{M_2}(K\cdot p^{M_1*}F).
\end{equation}
For the particular case of Riemann integral as a pushforward map, the Fourier-Mukai transform boils down to the familiar Fourier transform.

Another choice of pushforward map, which will allow us to find the dual R-R field strengths, is the fibrewise integration. Let us consider the case when the fibre bundle is the trivial bundle. Namely, this means that we have a projection $p:M\times\widehat{T}^n\times T^n\rightarrow M\times\widehat{T}^n$ from the fibre to the base and the fibrewise integration maps differential forms on $M\times\widehat{T}^n\times T^n$ to differential forms on $M\times\widehat{T}^n$ by means of the following rule:
\begin{equation}\label{eq:fibreint}
p_*\left(f(x,\hat{t}_i,t_i)\,\omega\wedge\dd t_{i_1}\wedge\ldots\wedge\dd t_{i_r}\right)\mapsto
\begin{dcases}
0&\text{if }r<n\\
\omega\int_{T^n}f(x,\hat{t}_i,t_i)\dd t_1\ldots\dd t_n&\text{if }r=n,
\end{dcases}
\end{equation}
where $x$ is a point from the manifold $M$, $\omega$ is a differential form on $M\times\widehat{T}^n$, and $f(x,\hat{t}_i,t_i)$ is some function of all coordinates.

The transformation rules of the R-R field strengths under N-AT-D have been worked out recently using Fourier-Mukai transform for the case of backgrounds possessing $SU(2)$ symmetry acting without isotropy. Our aim is to use this procedure for the PW geometry, which has the same symmetry. The concrete formula obtained in \cite{Gevorgyan:2013xka} and applied for the principal chiral model (note the plus sign in the curvature term \eqref{eq:curvature}, which is a consequence of the use of left-invariant one-forms instead of right-invariant ones) states:
\begin{equation}\label{eq:f-m}
\widehat{\mathcal{G}}=\int_G\mathcal{G}\wedge e^{\widehat{B}-B-dx^a\wedge L^a+\frac{1}{2}x^af_{bc}^aL^b\wedge L^c},
\end{equation}
where $\mathcal{G}=\sum_p\mathcal{G}_p$ is the sum of the gauge invariant R-R $p$-form field strengths and $\widehat{\mathcal{G}}$ is the sum of the dual ones; the index $p$ takes even values for type IIA and odd values for type IIB supergravities. Let us note that a factor of $L^2$, multiplying the curvature term, is necessary to ensure the correct dimensionality of the exponent. This factor is omitted here for the sake of brevity, but it will be restored in the final expressions. We will work in the co-frame of left-invariant one-forms $L^a$, in which $\mathcal{G}$ can be always represented as sum of differential forms that do not contain any $L^a$, differential forms that contain one $L^a$, differential forms that contain wedge product of two $L^a$'s, and differential forms that contain wedge products of three $L^a$'s:
\begin{equation}\label{eq:sumR-R}
\mathcal{G}=\mathcal{G}^{(0)}+\mathcal{G}_a^{(1)}\wedge L^a+\frac{1}{2}\mathcal{G}_{ab}^{(2)}\wedge L^a\wedge L^b+\mathcal{G}^{(3)}\wedge L^1\wedge L^2\wedge L^3.
\end{equation}
All considerations heretofore were quite general. Now we will specify them for the particular case of PW geometry ($L^a\equiv\sigma_a$). To do so, we sum all non-trivial R-R field strengths supported by the PW geometry as in \eqref{eq:sumR-R}. By making use of appendix \ref{sec:PW-forms}, we then determine the explicit form of all non-zero components of the differential forms $\mathcal{G}^{(0)}$, $\mathcal{G}_a^{(1)}$, $\mathcal{G}_{ab}^{(2)}$, $\mathcal{G}^{(3)}$:
\begin{align}\label{eq:R-R-field-strengths}
&\mathcal{G}^{(0)}=F_5^{(1)}\text{vol}(\text{AdS}_5),\quad\mathcal{G}^{(3)}=F_5^{(2)}\dd\theta\wedge\dd\phi,\quad\mathcal{G}_1^{(1)}=F_3^{(3)}\dd\theta\wedge\dd\phi+ F_7^{(2)}\text{vol}(\text{AdS}_5)\wedge\dd\theta,\nonumber\\
&\mathcal{G}_2^{(1)}=F_7^{(3)}\text{vol}(\text{AdS}_5)\wedge\dd\phi,\quad\mathcal{G}_{13}^{(2)}=F_3^{(1)}\dd\theta,\quad\mathcal{G}_{23}^{(2)}=F_3^{(2)}\dd\phi+F_7^{(1)}\text{vol}(\text{AdS}_5),
\end{align}
where we have introduced for brevity the notation $\text{vol}(\text{AdS}_5)=\dd\tau\wedge\dd\rho\wedge\dd\phi_1\wedge \dd\phi_2\wedge \dd\phi_3$. Using the already calculated in section \ref{sec:N-AT-D-PW} components of the dual $B$-field, the gauge invariant flux $\mathcal{F}$ in \eqref{eq:f-m} can be represented as a sum of the following two-forms:
\begin{align}
A^{(2,0)}&=\widehat{B}_{x_ax_b}\dd x_a\wedge\dd x_b,& A^{(1,0)}&=\widehat{B}_{\phi x_a}\dd\phi\wedge\dd x_a,&\nonumber\\*
A^{(1,1)}&=-\dd x_a\wedge\sigma_a,& A^{(0,0)}&=\widehat{B}_{\theta\phi}\dd\theta\wedge\dd\phi,&\nonumber\\
A^{(0,2)}&=-\epsilon_{abc}x_a\,\sigma_b\wedge\sigma_c-B_{\sigma_3\sigma_2}\,\sigma_3\wedge\sigma_2,& A^{(0,1)}&=-B_{\theta\sigma_1}\dd\theta\wedge\sigma_1-B_{\phi\sigma_2}\dd\phi\wedge\sigma_2.&
\end{align}
The gist of the above notations is the following. The first and the second digits in the superscript parentheses give information about the degree of the corresponding two-form in $dx_a$ and $\sigma_a$ respectively. In order to perform the fibrewise integration without missing any terms out, one should expand the exponent in \eqref{eq:f-m} up to fourth order in the gauge invariant flux,
\begin{equation}
e^{\mathcal{F}}=1+\mathcal{F}+\mathcal{F}^2+\mathcal{F}^3+\mathcal{F}^4+\mathcal{O}(\mathcal{F}^5),
\end{equation}
where $\mathcal{F}^n$ means the wedge product of $n$ differential forms $\mathcal{F}$. Having in mind that one should keep at most three-forms in $dx_a$ and $\sigma_a$ and also that wedge products of repeating differentials vanish, we sifted out all terms that will make contribution to the dual R-R field strengths in appendix \ref{sec:AppB}. The fibrewise integration \eqref{eq:fibreint} gives non-zero result only for differential forms that contain the three-form $\text{vol}(SU(2))=\sigma_1\wedge\sigma_2\wedge \sigma_3$ and this result is $\int_G\text{vol}(SU(2))=1$. Hence one should choose from expressions \eqref{eq:F1}--\eqref{eq:F4} all differential forms that contain wedge products of three, two, one, and zero $\sigma_a$'s, wedge them by $\mathcal{G}^{(0)}$, $\mathcal{G}_a^{(1)}$, $\mathcal{G}_{ab}^{(2)}$, and $\mathcal{G}^{(3)}$ respectively, and integrate the product along the fibre. The result of this laborious task is given in appendix \ref{sec:AppB} by equations \eqref{eq:G0}--\eqref{eq:G3}. We also have to restore the correct dimensionality of the field strengths by multiplying by $L^2$ wherever needed. Finally, we collect the differential forms by their degree in order to obtain the N-AT-D R-R field strengths of the dual type IIA theory. The dual two-form field strength is
\begin{align}\label{eq:dualF2}
\widehat{F}_2&=\left[B_{\phi\sigma_2}F_3^{(1)}+B_{\theta\sigma_1}F_3^{(2)}-(2L^2x_1+B_{\sigma_2\sigma_3})F_3^{(3)}+ F_5^{(2)}\right]\dd\theta\wedge\dd\phi\nonumber\\
&\phantom{{}={}}+L^2F_3^{(1)}\dd\theta\wedge\dd x_2-L^2F_3^{(2)}\dd\phi\wedge\dd x_1.
\end{align}
The four-form field strength has legs only in the space dual to the squashed five-sphere of the original PW geometry
\begin{equation}\label{eq:dualF4}
\widehat{F}_4=\widehat{F}_4^{(1)}\dd\theta\wedge\dd\phi\wedge\dd x_a\wedge\dd x_b+\left(\widehat{F}_4^{(2)} \dd\theta+\widehat{F}_4^{(3)}\dd\phi\right)\wedge\dd x_1\wedge\dd x_2\wedge\dd x_3,
\end{equation}
where
\begin{align}
\widehat{F}_4^{(1)}&=\widehat{B}_{x_ax_b}\left[B_{\phi\sigma_2}F_3^{(1)}+B_{\theta\sigma_1}F_3^{(2)}-(2L^2x_1+ B_{\sigma_2\sigma_3})F_3^{(3)}+F_5^{(2)}\right]+L^2\widehat{B}_{\phi x_a}\delta_b^2F_3^{(1)}\nonumber\\
&-\frac{L^4}{2}\epsilon_{1ab}F_3^{(3)},\quad\widehat{F}_4^{(2)}=2L^2\widehat{B}_{x_3x_1}F_3^{(1)},\quad\widehat{F}_4^{(3)}=2L^2\widehat{B}_{x_3x_2}F_3^{(2)}.
\end{align}
The dual six-form field strength has structure with legs in the whole non-abelian T-dual geometry
\begin{equation}\label{eq:dualF6}
\widehat{F}_6=\text{vol}(\text{AdS}_5)\wedge\left(\widehat{F}_6^{(1)}\dd\theta+\widehat{F}_6^{(2)}\dd\phi+ \widehat{F}_6^{(3)}\dd x_a\right),
\end{equation}
with
\begin{align}
\widehat{F}_6^{(1)}&=(2L^2x_1+B_{\sigma_2\sigma_3})\left(B_{\theta\sigma_1}F_5^{(1)}-F_7^{(2)}\right)-B_{\theta\sigma_1} F_7^{(1)},\nonumber\\
\widehat{F}_6^{(2)}&=2L^2x_2\left(B_{\phi\sigma_2}F_5^{(1)}-F_7^{(3)}\right),\quad\widehat{F}_6^{(3)}=L^2\left(2L^2x_a+ \delta_a^1B_{\sigma_2\sigma_3}\right)F_5^{(1)}-L^2\delta_a^1F_7^{(1)}.
\end{align}
The dual eight-form field strength is much more complicated
\begin{align}\label{eq:dualF8}
\widehat{F}_8=\text{vol}(\text{AdS}_5)\wedge&\left(\widehat{F}_8^{(1)}\dd x_1\wedge\dd x_2\wedge\dd x_3+ \widehat{F}_8^{(2)}\dd\theta\wedge\dd x_a\wedge\dd x_b\right.\nonumber\\
&\left.+\widehat{F}_8^{(3)}\dd\phi\wedge\dd x_a\wedge\dd x_b+\widehat{F}_8^{(4)}\dd\theta\wedge\dd\phi \wedge\dd x_a\right),
\end{align}
where
\begin{align}
\widehat{F}_8^{(1)}&=\left[L^6+2L^2\left(L^2\epsilon_{abc}\widehat{B}_{x_ax_b}x_c+\widehat{B}_{x_2x_3} B_{\sigma_2\sigma_3}\right)\right]F_5^{(1)}+2L^2\widehat{B}_{x_3x_2}F_7^{(1)},\nonumber\\
\widehat{F}_8^{(2)}&=L^4\delta_a^2\delta_b^3B_{\theta\sigma_1}F_5^{(1)}+\widehat{B}_{x_ax_b}\left[\left(2L^2x_1+ B_{\sigma_2\sigma_3}\right)\left(B_{\theta\sigma_1}F_5^{(1)}-F_7^{(2)}\right)-B_{\theta\sigma_1}F_7^{(1)}\right]-\frac{L^4}{2}\epsilon_{1ab} F_7^{(2)},\nonumber\\
\widehat{F}_8^{(3)}&=\left[L^2\widehat{B}_{\phi x_a}\left(2L^2x_b+\delta_b^1B_{\sigma_2\sigma_3}\right)+ L^2\left(2x_2\widehat{B}_{x_ax_b}-L^2\delta_a^1\delta_b^3\right)B_{\phi\sigma_2}\right]F_5^{(1)}-L^2\widehat{B}_{\phi x_a}\delta_b^1F_7^{(1)}\nonumber\\
&\phantom{{}={}}-\left(\frac{L^4}{2}\epsilon_{2ab}+ 2L^2x_2\widehat{B}_{x_ax_b}\right)F_7^{(3)},\nonumber\\
\widehat{F}_8^{(4)}&=\left[L^2\widehat{B}_{\theta\phi}\left(2L^2x_a+\delta_a^1B_{\sigma_2\sigma_3}\right)+L^2\delta_a^3 B_{\theta\sigma_1}B_{\phi\sigma_2}\right]F_5^{(1)}-\left(\widehat{B}_{\phi x_a}B_{\theta\sigma_1}+L^2\delta_a^1 \widehat{B}_{\theta\phi}\right)F_7^{(1)}\nonumber\\
&\phantom{{}={}}+\widehat{B}_{\phi x_a}\left(2L^2x_1+B_{\sigma_2\sigma_3}\right)\left(B_{\theta\sigma_1}F_5^{(1)}-F_7^{(2)}\right)-L^2\delta_a^3\left(B_{\phi\sigma_2}F_7^{(2)}+B_{\theta\sigma_1}F_7^{(3)}\right).
\end{align}
The dual ten-form field-strength is given by
\begin{equation}\label{eq:dualF10}
\widehat{F}_{10}=\widehat{F}_{10}^{(1)}\text{vol}(\text{AdS}_5)\wedge\dd\theta\wedge\dd\phi\wedge\dd x_1\wedge \dd x_2\wedge\dd x_3,
\end{equation}
with component
\begin{align}\label{eq:dualF101}
&\hspace{-5pt}\widehat{F}_{10}^{(1)}=\left[L^4\widehat{B}_{\phi x_1}B_{\theta\sigma_1}+L^6\widehat{B}_{\theta\phi}+ \epsilon_{abc}\widehat{B}_{\phi x_a}\widehat{B}_{x_bx_c}B_{\theta\sigma_1}\left(2L^2x_1+B_{\sigma_2\sigma_3}\right)+ 2L^2\widehat{B}_{x_1x_2}B_{\theta\sigma_1}B_{\phi\sigma_2}\right.\nonumber\\
&\hspace{-7pt}\left.+2L^2\widehat{B}_{\theta\phi}\left(L^2\epsilon_{abc}\widehat{B}_{x_ax_b}x_c+\widehat{B}_{x_2x_3} B_{\sigma_2\sigma_3}\right)\right]F_5^{(1)}-\left[\epsilon_{abc}\widehat{B}_{\phi x_a}\widehat{B}_{x_bx_c}B_{\theta\sigma_1}+ L^2\epsilon_{1ab}\widehat{B}_{x_ax_b}\widehat{B}_{\theta\phi}\right]F_7^{(1)}\nonumber\\
&\hspace{-7pt}-\left[L^4\widehat{B}_{\phi x_1}+\left(B_{\sigma_2\sigma_3}-L^2x_1\right)\epsilon_{abc}\widehat{B}_{\phi x_a}\widehat{B}_{x_bx_c}\right]F_7^{(2)}-L^2\epsilon_{3ab}\widehat{B}_{x_ax_b}\left(B_{\phi\sigma_2}F_7^{(2)}+ B_{\theta\sigma_1}F_7^{(3)}\right).
\end{align}
To finalise this section, let us make a noteworthy comment. The fibrewise integration in the Fourier-Mukai transform commutes with the exterior derivative. Consequently, if the R-R field strengths $\mathcal{G}$ are closed under the action of $\dd_H=\dd-H\wedge$ with $H=\dd B$, the non-abelian T-dual field strengths $\widehat{\mathcal{G}}$ are also closed under the action of $\dd_{\widehat{H}}=\dd-\widehat{H}\wedge$ with $\widehat{H}=\dd\widehat{B}$ \cite{Bouwknegt:2003vb}. This means that if the R-R field strengths $\mathcal{G}$ obey Bianchi identities, so do the T-dual field strengths $\widehat{\mathcal{G}}$. Since the gauge invariance is guaranteed by the Fourier-Mukai transform as well, one can infer that formula \eqref{eq:f-m} is an isomorphism.

\section{Kosmann derivative and supersymmetry conservation}\label{sec:susy}

Another intriguing question is, what amount of supersymmetry does the newly obtained N-AT-D theory possess? In \cite{Sfetsos:2010uq,Itsios:2012dc} the authors observed that the supersymmetry is conserved under T-duality when the Kosmann spinorial Lie derivative vanishes. Additionally, it has been shown \cite{Kelekci:2014ima} that this is equivalent to the condition the Killing spinor to be independent of the directions in which the T-duality (abelian or not) is performed. Here we will repeat the Kosmann derivative analysis of \cite{Kelekci:2014ima} for the following $(7,3)$-split metric ansatz:
\begin{equation}\label{eq:metricansatz}
	\dd s_{10}^2=\dd s_7^2+\sum_{a=1}^3e^{2C_a}(\sigma_a+\mathcal{A}^a)^2,
\end{equation}
which is perfectly consistent with the PW solution in its IR point. Here $\sigma_a$ are the same left-invariant one-forms as defined in formula \eqref{eq:left-inv-forms}, $\mathcal{A}^a$ are $SU(2)$-valued one-forms, and $C_a$ are some scalar warp functions\footnote{Both $\mathcal{A}^a$ and $C_a$ depend on the transverse coordinates.}. Explicitly, the PW metric in the above notations takes the form:
\begin{align}
	\dd s_7^2&=\dd s_{1,4}^2(\text{IR})+\frac{2}{3}L^2\Omega^2\left[\dd\theta^2
	+\frac{2\left(7-6\cos2\theta+3\cos^22\theta\right)}{3\left(3-\cos2\theta\right)^2}\right],\nonumber\\
	e^{2C_1}&=e^{2C_2}=\frac{2}{3}L^2\Omega^2\frac{4\cos^2\theta}{3-\cos2\theta},\quad
	e^{2C_3}=\frac{2}{3}L^2\Omega^2\frac{8\cos^2\theta\left(5-\cos2\theta\right)}
	{3\left(3-\cos2\theta\right)^2},\nonumber\\
	\mathcal{A}^1&=\mathcal{A}^2=0,\quad\mathcal{A}^3=\frac{2}{5-\cos2\theta}\dd\phi.
\end{align}
When all functions $C_a$ are different, the metric \eqref{eq:metricansatz} has right-acting $SU(2)$ isometry. If two of the $C_a$'s are equal, the isometry is enhanced to $SU(2)\times U(1)$, and when the three scalars are equal the metric possesses the $SO(4)$ symmetry of a round $S^3$. This is in full agreement with the $SU(2)\times U(1)_\phi$ isometry of the IR PW metric, since in our case $C_1=C_2$. Introducing natural orthonormal frame in \eqref{eq:metricansatz},
\begin{equation}\label{eq:frame}
	e^\mu=\bar{e}^\mu,\quad e^a=e^{C_a}\left(\sigma_a+\mathcal{A}^a\right),
\end{equation}
one can readily compute the spin connection
\begin{align}
	\omega_{\phantom{1}2}^1&=\frac{1}{2}e^{-C_1-C_2-C_3}\left(e^{2C_1}+e^{2C_2}-e^{2C_3}\right)e^3
	-\frac{1}{2}e^{-C_1-C_2}\left(e^{2C_1}+e^{2C_2}\right)\mathcal{A}^3,\nonumber\\
	\omega_{\phantom{1}\mu}^1&=\partial_\mu C_1e^1-\frac{1}{2}e^{-C_1-C_2}\left(e^{2C_2}-e^{2C_1}\right)
	\mathcal{A}_\mu^3e^2+\frac{1}{2}e^{-C_1-C_3}\left(e^{2C_3}-e^{2C_1}\right)\mathcal{A}_\mu^2e^3
	+\frac{1}{2}e^{C_1}\mathcal{F}_{\phantom{1}\mu\rho}^1e^\rho,\nonumber\\
	\omega_{\phantom{\mu}\nu}^\mu&=\bar{\omega}_{\phantom{\mu}\nu}^\mu-\sum_a\frac{1}{2}e^{C_a}
	\mathcal{F}_{\phantom{a\mu}\nu}^{a\mu}e^a,
\end{align}
where the cyclic terms in $a=1,2,3$ are inferred and $\mathcal{F}^a=\dd\mathcal{A}^a+\frac{1}{2}\epsilon_{abc}
\mathcal{A}^b\wedge\mathcal{A}^c$ are the non-abelian field strengths corresponding to the $SU(2)$ gauge fields $\mathcal{A}^a$.

Following \cite{Kelekci:2014ima}, the next step is to prove that the Kosmann derivative with respect to the right-invariant Killing vector fields $K_a$, applied to the Killing spinor $\eta$ of the PW background, vanishes if and only if the spinor does not depend on the Killing directions. As we mentioned earlier, this is equivalent to supersymmetry preservation under N-AT-D if the original Killing spinor does not depend on the directions on which the T-duality is performed. Firstly, we need the right-invariant vector fields, which are dual to the right-invariant one-forms parameterising our $SU(2)$ isometry. Additionally, the right-invariant one-forms must be consistent with the parameterisation of $SU(2)$ we have chosen by the definition of the left-invariant one-forms \eqref{eq:left-inv-forms}. A straightforward computation gives the following result:
\begin{align}
	K_1&=-2\left(\cot\alpha\cos\gamma\,\partial_\gamma+\sin\gamma\,\partial_\alpha
	-\frac{\cos\gamma}{\sin\alpha}\partial_\beta\right),\nonumber\allowdisplaybreaks[0]\\
	K_2&=2\left(\cot\alpha\sin\gamma\,\partial_\gamma-\cos\gamma\,\partial_\alpha
	-\frac{\sin\gamma}{\sin\alpha}\partial_\beta\right),\nonumber\allowdisplaybreaks[0]\\
	K_3&=2\partial_\gamma.
\end{align}
The right-invariant vectors themselves can be decomposed in the covariant derivatives $\nabla_a$ with respect to the left-invariant one-forms $\sigma_a$,
\begin{align}\label{eq:decomposition}
	K_1^a\nabla_a\eta&=\left[e^{C_1}\left(\cos\alpha\cos\beta\cos\gamma-\sin\beta\sin\gamma\right)\nabla_1
	\right.\nonumber\\
	&\phantom{{}={}}\left.+e^{C_2}\left(\cos\alpha\sin\beta\cos\gamma+\cos\beta\sin\gamma\right)\nabla_2
	+e^{C_3}\sin\alpha\cos\gamma\nabla_3\right]\eta,\nonumber\\
	K_2^a\nabla_a\eta&=\left[-e^{C_1}\left(\cos\alpha\cos\beta\sin\gamma+\sin\beta\cos\gamma\right)\nabla_1
	\right.\nonumber\\
	&\phantom{{}={}}\left.-e^{C_2}\left(\cos\alpha\sin\beta\sin\gamma-\cos\beta\cos\gamma\right)\nabla_2
	-e^{C_3}\sin\alpha\sin\gamma\nabla_3\right]\eta,\nonumber\\
	K_3^a\nabla_a\eta&=\left[-e^{C_1}\sin\alpha\cos\beta\nabla_1-e^{C_2}\sin\alpha\sin\beta\nabla_2
	+e^{C_3}\cos\alpha\nabla_3\right]\eta.
\end{align}
The above expressions are exactly the first terms in the corresponding Lie derivatives of the spinor $\eta$ with respect to the Killing vector fields $K_i$, $i=1,2,3$. Namely, the Kosmann derivative of a spinor field $\eta$ is defined as \cite{Kosmann1972}
\begin{equation}\label{eq:KosmannDerivative}
	\pounds_{K_i}\eta=K_i^a\nabla_a\eta+\frac{1}{8}\left(\dd K_i\right)_{ab}\Gamma^{ab}\eta,
\end{equation}
where $\nabla_a\equiv\partial_a+\frac{1}{4}\omega_{abc}\Gamma^{bc}$. In order to compute the second term in \eqref{eq:KosmannDerivative}, we have to compute $\dd K_i$ and make use of the following identities
\begin{align}
	\dd\alpha&=2\sin\beta\left(e^{-C_1}e^1-A^1\right)-2\cos\beta\left(e^{-C_2}e^2-A^2\right),\nonumber\\
	\dd\beta&=2\cos\beta\cot\alpha\left(e^{-C_1}e^1-A^1\right)
	+2\sin\beta\cot\alpha\left(e^{-C_2}e^2-A^2\right)+2\left(e^{-C_3}e^3-A^3\right),\nonumber\\
	\dd\gamma&=-2\frac{\cos\beta}{\sin\alpha}\left(e^{-C_1}e^1-A^1\right)
	-2\frac{\sin\beta}{\sin\alpha}\left(e^{-C_2}e^2-A^2\right),
\end{align}
which can be easily derived by inverting the expressions for the orthonormal frame $e^a$. Now we are in position to show that formula \eqref{eq:KosmannDerivative} can be recast in the form
\begin{equation}
	\pounds_{K_i}\eta=K_i^aP_a\eta,\quad i=1,2,3,
\end{equation}
where $K_i^a$ are trigonometric functions of $\alpha$, $\beta$, and $\gamma$ that can be read off from \eqref{eq:decomposition}, and $P_a\eta$ are expressions encoding the left-invariant forms alone \eqref{eq:left-inv-forms}. Now one can readily conclude that
\begin{equation}
	\pounds_{K_i}\eta=0\Leftrightarrow P_a\eta=0\Leftrightarrow
	\eta~\text{is constant w.r.t.}~(\alpha,\beta,\gamma).
\end{equation}
In \cite{Pilch:2000ej} Pilch and Warner found that the spinor does not depend on the $SU(2)$ directions. One should be careful, however, because the last statement is frame-dependent. The two frames---\eqref{eq:frame} and the frame in \cite{Pilch:2000ej}---are not the same, but they differ only by rotation that will not induce any $SU(2)$ coordinate dependence in the spinor. Therefore the amount of supersymmetry of the original PW background should stay unaffected under the N-AT-D. Since the IR-fixed-point Pilch-Warner solution has $\mathcal{N}=1$ supersymmetry, we conclude that its non-abelian T-dual geometry also possesses $\mathcal{N}=1$ supersymmetry.

\section{pp-wave limits of the T-dual geometry}\label{sec:pp-wave}

In this section we will consider two different light-like (null) geodesics of the geometry we have obtained applying N-AT-D. According to Penrose \cite{Penrose1976}, for any spacetime exists certain limit in which the geometry becomes of a plane wave type. This limit has been further generalised for supergravity solutions in string theory by G\"uven \cite{Gueven:2000ru}. The idea is to zoom into the geometry seen by a particle moving very fast on a null geodesic. From supergravity point of view, the Penrose-G\"uven limit results in pp-wave geometry, which is $\alpha'$-exact, possesses globally defined null Killing vector field, and is still solution of Einstein's supergravity equations of motion. The procedure of taking the limit consists of finding a light-like geodesic, introducing light-cone coordinates, and blowing up a neighbourhood of the null geodesic using the property that the Einstein-Hilbert action is homogeneous under constant scalings of the metric.

\subsection{The $\theta=0$ geodesic}

The warp factor $\Omega^2$ that deforms the original geometry depends on angle $\theta$, which parametrises the five-sphere. Therefore we are forced to choose $\theta$ be some constant. Any choice of constant, however, picks different geodesic, which must be consistent with the Penrose limit and consequently the blow up around the selected geodesic must give finite results for the metric and all fields. Then the simplest null geodesic one can select is: $\theta=0$, $\rho=0$, $x_1=0$, $x_2=0$. The metric, restricted onto this geodesic, takes the form
\begin{equation}
	\dd s^2=-\frac{2^{5/6}L^2}{\sqrt{3}}\dd\tau^2
	+\frac{9\sqrt{3}L^2}{16\times2^{5/6}}\dd x_3^2.
\end{equation}
Hence one can readily introduce light-cone coordinates given by the following linear coordinate transformations:
\begin{equation}\label{eq:light-cone}
	x^+=\frac{2^{5/12}}{3^{1/4}}\tau+\frac{3\times3^{1/4}}{4\times2^{5/12}}x_3,\quad
	x^-=\frac{2^{5/12}}{3^{1/4}}\tau-\frac{3\times3^{1/4}}{4\times2^{5/12}}x_3.
\end{equation}
Having introduced the light-cone coordinates $x^+$ and $x^-$, the Penrose-G\"uven limit consists of multiplying the metric by overall factor of $\Lambda^{-2}$, where $\Lambda$ is some parameter. Next one has to appropriately rescale all coordinate variables (except for $x^+$, which is the affine parameter on the geodesic) that define the light-like geodesic using the same parameter $\Lambda$:
\begin{align}\label{eq:resc}
	x^+&=\frac{2^{5/12}}{3^{1/4}}u,\quad
	x^-=\frac{2\times3^{1/4}}{2^{5/12}}\Lambda^2v,\quad
	\rho=\frac{3^{1/4}}{2^{5/12}}\Lambda r,\nonumber\\
	\theta&=\frac{3\times3^{1/4}}{2^{11/12}}\Lambda\alpha,\quad
	x_1=\frac{2\times2^{5/12}}{3^{3/4}}\Lambda y_1,\quad
	x_2=\frac{2\times2^{5/12}}{3^{3/4}}\Lambda y_2.
\end{align}
All the constants in the above formula are chosen for the sake of clearing out the final result. Then blowing up a neighbourhood of the selected geodesic corresponds to the limit $\Lambda\rightarrow0$. The result of this limit for our metric is
\begin{align}\label{eq:ds2pp-wave1}
	\dd s_\text{pp-wave}^2=&
	-2L^2\dd u\dd v+\frac{3L^2}{2\left(1+u^2\right)}
	\left[u\left(y_1+\alpha\right)\dd y_1+uy_2\dd y_2\right.\nonumber\\
	&\left.-\left(u\left(y_1+\alpha\right)-y_2\right)\dd\alpha
	-2\alpha\left(y_1+uy_2+\alpha\right)\dd\phi\right]\dd u\nonumber\\
	&-\frac{L^2}{16\left(1+u^2\right)}\left[4r^2\left(1+u^2\right)
	+9\left(y_1^2+y_2^2+2y_1\alpha-\left(2+3u^2\right)\alpha^2\right)\right]\dd u^2\nonumber\\
	&+\frac{L^2}{1+u^2}\left[\dd y_1^2+\dd y_2^2+\left(4+3u^2\right)\dd\alpha^2
	+\left(y_1^2+y_2^2-2y_1\alpha+\left(4+3u^2\right)\alpha^2\right)\dd\phi^2\right]\nonumber\\
	&+\frac{2L^2}{1+u^2}\left[y_2\dd\alpha\dd\phi-\dd\alpha\dd y_1-u\dd\alpha\dd y_2
	-2\alpha\dd\phi\dd y_2+2u\alpha\dd\phi\dd y_1\right]\nonumber\\
	&+L^2\left[\dd r^2+r^2\left(\dd\phi_1^2+\sin^2\phi_1\left(\dd\phi_2^2
	+\sin^2\phi_2\dd\phi_3^2\right)\right)\right].
\end{align}
The obtained metric is of the most general pp-wave type. As expected, the AdS part $(r,\phi_1,\phi_2,\phi_3)$ represents a four-dimensional flat subspace written as metric on a three-sphere via pullback of the Euclidean metric on $\mathbb{R}^4$. The rest of the metric \eqref{eq:ds2pp-wave1}, namely the space spanned by the coordinates $(y_1,y_2,\alpha,\phi)$, is tied to the affine parameter $u$ on the geodesic and has non-zero curvature.

The next step is to apply the described above procedure for the dual NS two-form $\widehat{B}$. However, if one proceed straightforwardly, they would encounter infinity that originates from the coefficient $\widehat{B}_{\phi x_3}$ in front of $\dd\phi\wedge\dd x_3$. This is actually a gauge feature that can be cured by a conveniently chosen gauge transformation on the $\widehat{B}$-field. One can easily check that the correct gauge transformation, which removes the infinity, is $-\frac{L}{2}\dd\phi\wedge\dd x_3$. Afterwards, the Penrose-G\"uven limit of the already gauge-transformed Kalb-Ramond field gives the finite result:
\begin{align}
	\tilde{B}=&\frac{L^2}{4\left(1+u^2\right)}\left[
	4u\left(y_1-\alpha\right)\dd\alpha\wedge\dd\phi-3y_2\dd u\wedge\dd y_1
	+3\left(y_1+\alpha\right)\dd u\wedge\dd y_2\right.\nonumber\\
	&\left.+3\left(y_1^2+y_2^2+2y_1\alpha+\left(2+u^2\right)\alpha^2\right)
	\dd u\wedge\dd\phi-4u\dd y_1\wedge\dd y_2\right.\nonumber\\
	&\left.-4\left(y_2+u\left(y_1+\alpha\right)\right)\dd y_1\wedge\dd\phi
	+4\left(y_1-uy_2+\alpha\right)\dd y_2\wedge\dd\phi\right].
\end{align}
The same limit should be calculated for the dual R-R field strengths as well. We need to introduce the same light-cone coordinates \eqref{eq:light-cone} in equations \eqref{eq:dualF2}--\eqref{eq:dualF101}, rescale the coordinates as in \eqref{eq:resc}, multiply every differential form by $\Lambda^{-p}$, where $p$ is the degree of the corresponding differential form, and lastly take the limit $\Lambda\rightarrow0$. The result for the pp-wave limit of the dual $\widehat{F}_2$ field strength \eqref{eq:dualF2} appears to be very simple,
\begin{equation}
	\tilde{F}_2=\frac{16\times2^{5/6}L^4}{3\sqrt{3}}\left[
	\alpha\dd y_1\wedge\dd\phi-\dd y_2\wedge\dd\alpha
	+\left(y_1-6\alpha\right)\dd\alpha\wedge\dd\phi\right].
\end{equation}
The limit of the non-abelian T-dual $\widehat{F}_4$ field strength \eqref{eq:dualF4} is
\begin{align}
	\tilde{F}_4=&\frac{4\times2^{5/6}L^6}{3\sqrt{3}\left(1+u^2\right)}\left[
	-6y_2\dd u\wedge\dd y_1\wedge\dd y_2\wedge\dd\alpha
	+6\alpha\left(y_1+\alpha\right)\dd u\wedge\dd y_1\wedge\dd y_2\wedge\dd\phi\right.\nonumber\allowdisplaybreaks[0]\\
	&\left.+6y_2\left(y_1-6\alpha\right)\dd u\wedge\dd y_1\wedge\dd\alpha\wedge\dd\phi
	-4\left(y_2-uy_1+13u\alpha\right)\dd y_1\wedge\dd y_2\wedge\dd\alpha\wedge\dd\phi\right.\nonumber
	\allowdisplaybreaks[0]\\
	&\left.+3\left(-y_1^2+y_2^2+12y_1\alpha+\left(8-5u^2\right)\alpha^2\right)
	\dd u\wedge\dd y_2\wedge\dd\alpha\wedge\dd\phi\right].
\end{align}
The Penrose-G\"uven limit of $\widehat{F}_6$ field strength \eqref{eq:dualF6} results in
\begin{align}
	\tilde{F}_6=&\frac{4\times2^{5/6}L^8}{9\sqrt{3}}r^3\sin^2\phi_1\sin\phi_2\left[
	16u\dd u\wedge\dd v\wedge\dd r+3\left(4y_1+9\alpha\right)\dd u\wedge\dd r\wedge\dd y_1\right.\nonumber\\
	&\left.+3\left(7y_1+12\alpha\right)\dd u\wedge\dd r\wedge\dd\alpha-12y_2\dd u\wedge\dd r\wedge\dd y_2
	+48y_2\alpha\dd u\wedge\dd r\wedge\dd\phi\right]\nonumber\\
	&\wedge\dd\phi_1\wedge\dd\phi_2\wedge\dd\phi_3.
\end{align}
The result for the limit of $\widehat{F}_8$ field strength \eqref{eq:dualF8} is
\begin{align}
	\tilde{F}_8=&-\frac{4\times2^{5/6}L^{10}}{9\sqrt{3}\left(1+u^2\right)}r^3\sin^2\phi_1\sin\phi_2
	\dd r\wedge\dd u\wedge\left[
	16\left(-1+u^2\right)\dd v\wedge\dd y_1\wedge\dd y_2\right.\nonumber\\
	&\left.+4\left(-4y_1+4uy_2+7\alpha+11u^2\alpha\right)\dd v\wedge\dd y_1\wedge\dd\phi
	-28\left(1+u^2\right)\dd v\wedge\dd y_2\wedge\dd\alpha\right.\nonumber\\
	&\left.-16\left(uy_1+y_2+u\alpha\right)\dd v\wedge\dd y_2\wedge\dd\phi
	+6u\left(7y_1+12\alpha\right)\dd y_1\wedge\dd y_2\wedge\dd\alpha\right.\nonumber\\
	&\left.+\left(12\left(y_1^2+y_2^2\right)+3\alpha\left(13y_1+27uy_2+9\alpha\right)\right)
	\dd y_1\wedge\dd y_2\wedge\dd\phi\right.\nonumber\\
	&\left.+3\left(y_1\left(3uy_1-7y_2\right)-2\alpha\left(7uy_1+6y_2\right)-21u\alpha^2\right)
	\dd y_1\wedge\dd\alpha\wedge\dd\phi\right.\nonumber\\
	&\left.+3\left(7y_1^2+4\alpha\left(-4uy_2+3\alpha\right)+y_1\left(-3uy_2+19\alpha\right)\right)
	\dd y_2\wedge\dd\alpha\wedge\dd\phi\right.\nonumber\\
	&\left.+4\left(y_1\left(7+3u^2\right)-2\alpha\left(5+3u^2\right)\right)
	\dd v\wedge\dd\alpha\wedge\dd\phi\right]\wedge\dd\phi_1\wedge\dd\phi_2\wedge\dd\phi_3.
\end{align}
Finally, the pp-wave limit of the dual $\widehat{F}_{10}$ field strength \eqref{eq:dualF10} gives
\begin{align}
	\tilde{F}_{10}=&-\frac{16\times2^{5/6}L^{12}}{9\sqrt{3}\left(1+u^2\right)^2}
	r^3\left[7y_2+u\left(\left(-2+6u^2\right)y_1+7uy_2+31\alpha+23u^2\alpha\right)\right]\nonumber\\
	&\times\sin^2\phi_1\sin\phi_2\dd u\wedge\dd v\wedge\dd r\wedge\dd y_1\wedge\dd y_2
	\wedge\dd\alpha\wedge\dd\phi\wedge\dd\phi_1\wedge\dd\phi_2\wedge\dd\phi_3.
\end{align}
In the next section we will slightly modify the limiting procedure, more specifically the rescaling of the coordinates, and will consider another geodesic. Even though the limit is not the same as prescribed by Penrose-G\"uven, it will lead again to pp-wave geometry with metric obtained directly in Rosen form.

\subsection{The $\theta=\pi/4$ geodesic}

Let us consider in this section another null geodesic, namely the one with $\theta=\pi/4$, $\rho=0$, $x_1=0$, $x_2=0$, $x_3=0$. The metric onto this geodesic takes the form
\begin{equation}
	\dd s^2=-2^{1/3}L^2\dd\tau^2+\frac{8\times2^{1/3}L^2}{27}\dd\phi^2.
\end{equation}
There is again a natural choice of light-cone coordinates. In contrast to the previous case, where we rescaled the coordinates asymmetrically, this time we will rescale them symmetrically with the same degree of the parameter $\Lambda$ for both light-cone coordinates. Note that this change does not supersede the basic idea underlying the Penrose-G\"uven limit, specifically this rescaling again represents zoom into the geometry around a selected geodesic. The light-cone coordinates are:
\begin{equation}
	x^+=\frac{1}{\Lambda}\left(2^{1/6}\tau+\frac{2\times2^{2/3}}{3\sqrt{3}}\phi\right),\quad
	x^-=\frac{1}{2\Lambda}\left(2^{1/6}\tau-\frac{2\times2^{2/3}}{3\sqrt{3}}\phi\right).
\end{equation}
The rest of the coordinates that define the null geodesic are rescaled as before by factor of $\Lambda$ and conveniently chosen constants in order to clear out the final result:
\begin{align}
	\rho&=2^{-1/6}\Lambda r,\quad
	\theta=\frac{\pi}{4}+\frac{3\times2^{1/3}}{\sqrt{16+9\times2^{1/6}}}\Lambda y^4,\nonumber\allowdisplaybreaks[0]\\
	x_1&=\frac{2\times2^{1/6}}{3}\Lambda y^1,\quad
	x_2=\frac{2\times2^{2/3}}{\sqrt{15}}\Lambda y^2,\quad
	x_3=\frac{4\times2^{1/6}}{3\sqrt{3}}\Lambda y^3.
\end{align}
Following the Penrose procedure we multiply the dual metric by overall factor of $\Lambda^{-2}$ and send $\Lambda\rightarrow0$. The result of the limit is
\begin{align}\label{eq:ds2pp-wave2}
	\dd s_\text{pp-wave}^2=
	L^2\bigg[-2\dd x^+\dd x^--\frac{1}{3\sqrt{2}}\left(\dd x^+-2\dd x^-\right)
	\left(\sqrt{5}\dd y^2+\dd y^3\right)-\frac{4}{a}\dd y^1\dd y^4\nonumber\\
	+\sum_{i=1}^4\dd y^i\dd y^i+\dd r^2+r^2\left(\dd\phi_1^2
	+\sin^2\phi_1\left(\dd\phi_2^2+\sin^2\phi_2\dd\phi_3^2\right)\right)\bigg],
\end{align}
where the constant $a=\sqrt{16+9\times2^{1/6}}$. The standard AdS part appears again as metric on $S^3$, while the subspace spanned by the coordinates $y^i$, $i=1,\ldots,4$, is much simpler compared to \eqref{eq:ds2pp-wave1}, although it is tied again to the light-cone coordinates $x^+$ and $x^-$. However, one can easily factorise the metric using the linear coordinate transformations,
\begin{equation}\label{eq:rosencoord}
	u=x^+-\frac{1}{3\sqrt{2}}\left(\sqrt{5}y^2+y^3\right),\quad
	v=x^-+\frac{1}{6\sqrt{2}}\left(\sqrt{5}y^2+y^3\right).
\end{equation}
Consequently, pushing forward the metric on $S^3$ to Euclidean metric on $\mathbb{R}^4$ with coordinates $y^i$, $i=5,\ldots,8$, brings the metric \eqref{eq:ds2pp-wave2} exactly to pp-wave metric in Rosen form,
\begin{equation}
	\dd s_\text{pp-wave}^2=L^2\left(-2\dd u\dd v+C_{ij}\dd y^i\dd y^j\right),
\end{equation}
where $i,j=1,\ldots,8$ and the constant matrix $C_{ij}$ has the form
\begin{equation}
	C_{ij}=
	\begin{pmatrix}
		1 & 0 & 0 & -2/a & 0\\
		0 & 13/18 & -\sqrt{5}/18 & 0 & 0\\
		0 & -\sqrt{5}/18 & 17/18 & 0 & 0\\
		-2/a & 0 & 0 & 1 & 0\\
		0 & 0 & 0 & 0 & \mathds{1}_{4\times4}
	\end{pmatrix}.
\end{equation}
What is left to conclude this section is to repeat all the steps described above for the NS-NS and R-R fields keeping in mind that we have to do one additional coordinate change \eqref{eq:rosencoord} in order to make all coordinates compatible. The pp-wave limit of the Kalb-Ramond $\widehat{B}$-field is
\begin{equation}
	\tilde{B}=-\frac{L^2}{30}\left[\left(\dd u-2\dd v\right)\wedge\left(
	\sqrt{10}\dd y^2-5\sqrt{2}\dd y^3\right)+2\sqrt{5}\dd y^2\wedge\dd y^3\right].
\end{equation}
Let us emphasise here that, contrary to the geodesic $\theta=0$, in this case the $\tilde{B}$-field is finite without making any gauge transformations. The limit of the dual $\widehat{F}_2$ field strength gives
\begin{align}
	\tilde{F}_2=&\frac{4\times2^{1/3}L^4}{405\sqrt{3}a}\left[\vphantom{\sqrt{5}}
	5\left(\dd u-2\dd v\right)\wedge\left(84\dd y^4-9a\dd y^1\right)\right.\nonumber\\
	&\left.+15\sqrt{2}a\dd y^1\wedge\left(\sqrt{5}\dd y^2+\dd y^3\right)
	+4\sqrt{2}\left(23\sqrt{5}\dd y^2+35\dd y^3\right)\wedge\dd y^4\right].
\end{align}
The result for the four-form $\widehat{F}_4$ field is
\begin{equation}
	\tilde{F}_4=\frac{8\times2^{1/3}L^6}{81\sqrt{15}a}\left(\dd u-2\dd v\right)\wedge\left[
	9a\dd y^1\wedge\dd y^2\wedge\dd y^3-46\dd y^2\wedge\dd y^3\wedge\dd y^4\right].
\end{equation}
The Penrose-G\"uven limit applied to the dual $\widehat{F}_6$ field strength results in
\begin{equation}
	\tilde{F}_6=\frac{4\times2^{5/6}L^8}{243a}r^3\sin^2\phi_1\sin\phi_2
	\left(\dd u+2\dd v\right)\wedge\dd r\wedge\left[19a\dd y^1-56\dd y^4
	\right]\wedge\dd\phi_1\wedge\dd\phi_2\wedge\dd\phi_3.
\end{equation}
The plane wave limit of the $\widehat{F}_8$ R-R field strength is
\begin{align}
	\tilde{F}_8=&\frac{8\times2^{1/3}L^{10}}{3645a}r^3\sin^2\phi_1\sin\phi_2\dd r\wedge
	\left[\dd u\wedge\dd v\wedge\left(38\sqrt{5}a\dd y^1\wedge\dd y^2\right.\right.\nonumber\allowdisplaybreaks[0]\\
	&\left.\left.+290a\dd y^1\wedge\dd y^3+112\sqrt{5}\dd y^2\wedge\dd y^4
	+940\dd y^3\wedge\dd y^4\right)\right.\nonumber\\
	&\left.+3\sqrt{10}\left(\dd u+2\dd v\right)\wedge\left(2a\dd y^1+23\dd y^4\right)
	\wedge\dd y^2\wedge\dd y^3\right]\wedge\dd\phi_1\wedge\dd\phi_2\wedge\dd\phi_3.
\end{align}
Finally, the pp-wave limit of the ten-form R-R field strength $\widehat{F}_{10}$ vanishes, $\tilde{F}_{10}=0$.

\section{Conclusion}\label{sec:conclusion}

In this paper we consider non-abelian T-duality of the well-known Pilch-Warner supergravity solution with non-trivial R-R fluxes. It is important to mention that the Pilch-Warner solution is a background with a certain $G$-structure\footnote{More information on the specific way the duality acts on the $G$-structure of the seed solutions can be found in \cite{Barranco:2013fza,Macpherson:2015tka}.}, which in this case is a global isometry group. This fact facilitates the calculation of its non-abelian T-dual counterpart. For this reason we adopt the procedure detailed in \cite{Gevorgyan:2013xka}, where one first gauges the isometry group, thus introducing some auxiliary gauge field variables and Lagrange multipliers. Integrating out the gauge fields yields a Lagrangian depending on the original variables and the Lagrange multipliers. After fixing the gauge the dual action is produced.

Following \cite{Gevorgyan:2013xka} we obtain the ten-dimensional non-abelian T-dual Pilch-Warner metric and the dual NS Kalb-Ramond B-field. The result for the dual metric is relatively complicated. It is a direct product of a warped $\text{AdS}_5$ space, with the same warp factor as in the original Pilch-Warner geometry, times a five-dimensional manifold $M_5$ with yet undetermined structure. On the other hand the non-vanishing components of the dual $B$-field retain simpler form.

We also obtain the type IIA dual field strengths of the original R-R Pilch-Warner fluxes via application of the Fourier-Mukai transform. The resulting expressions for the $\widehat{F}_2$ and $\widehat{F}_4$ forms are not so complicated with only few legs in the dual $M_5$ space. The resulting $\widehat{F}_6$, $\widehat{F}_8$, and $\widehat{F}_{10}$ dual field strengths have messy non-vanishing components and many legs in the full ten-dimensional space. Since the derived N-AT-D geometry of the IR PW supergravity solution is quite complicated and with not so clear structure, we have computed two different pp-wave limits. This immediately motivates one possible extension of the present work, i.e. construction of BMN operators at least for the case $\theta=\pi/4$. We leave this question to a future work. Another interesting direction in which the present work could be extended is to consider what happens with the R-R fields if one makes a large gauge transformation on the initial $B$-field or the dual $\widehat{B}$-field.

\section*{Acknowledgements}

We would like to thank Carlos Nunez, Niall Macpherson, Daniel Thompson, Eoin O Colgain, Leopoldo A. Pando Zayas, and Thiago Rocha for the valuable comments on the manuscript. RR thanks Kostya Zarembo for comments on the Killing spinor in Pilch-Warner background. This work was partially supported by the Bulgarian NSF grant DFNI T02/6. RR is supported in part by the Austrian Science Fund (FWF) project I 1030-N16. He also acknowledges the partial support from the SEENET-Ni\v{s} office and thanks for the opportunity to present part of these results there.

\begin{appendix}

\section{R-R field strengths of the original PW solution}\label{sec:PW-forms}

The 10-dimensional Pilch-Warner supergravity solution \cite{Pilch:2000fu} is equipped with non-trivial Ramond-Ramond (R-R) and Neveu-Schwarz (NS-NS) fluxes. Here we consider only the infrared critical point of the RG flow, where the solution describes warped $\text{AdS}_5$ times squashed $S^5$ space. In this case the dilaton and the axion are constant along the flow. Generally, all NS and R-R fluxes and their corresponding field strengths satisfy certain equations of motion \cite{PremKumar:2010as,Maxfield:2013wka}. The three-form R-R field strength is given by:
\begin{equation}
F_3=F_3^{(1)}\dd\theta\wedge\sigma_1\wedge\sigma_3+F_3^{(2)}\dd\phi\wedge\sigma_2\wedge\sigma_3+ F_3^{(3)}\dd\theta\wedge\dd\phi\wedge\sigma_1,
\end{equation}
where $\sigma_i$ are the $SU(2)$ left-invariant one-forms defined in section \ref{sec:N-AT-D-PW} and
\begin{align}
F_3^{(1)}&=\frac{64\times2^{1/3}L^2\cos^3\theta}{9(3-\cos2\theta)^2},\quad F_3^{(2)}=\frac{32\times2^{1/3}L^2\cos^2\theta\sin\theta}{9(3-\cos2\theta)},\nonumber\\
F_3^{(3)}&=\frac{4\times2^{1/3}L^2\cos\theta(11-20\cos2\theta+\cos4\theta)}{9(3-\cos2\theta)^2}.
\end{align}
The self-dual five-form field strength has the following form:
\begin{equation}
F_5=F_5^{(1)}\dd\tau\wedge\dd\rho\wedge\dd\phi_1\wedge\dd\phi_2\wedge\dd\phi_3+F_5^{(2)}\dd\theta\wedge\dd\phi \wedge\sigma_1\wedge\sigma_2\wedge\sigma_3,
\end{equation}
where the electric and magnetic parts are given by
\begin{equation}
F_5^{(1)}=-\frac{8}{3}2^{2/3}L^4\cosh\rho\sinh^3\rho\sin^2\phi_1\sin\phi_2,\quad F_5^{(2)}=-\frac{1024\times2^{2/3}L^4\cos^3\theta\sin\theta}{81(3-\cos2\theta)^2}.
\end{equation}
The seven-form field strength is Hodge dual to $F_3$, $F_7=\star F_3$. Its explicit form in the co-frame of left-invariant one-forms $\sigma_i$ is:
\begin{equation}
F_7=\text{vol}(\text{AdS}_5)\wedge\left(F_7^{(1)}\sigma_2\wedge\sigma_3+F_7^{(2)}\dd\theta\wedge\sigma_1+ F_7^{(3)}\dd\phi\wedge\sigma_2\right),
\end{equation}
where
\begin{align}
F_7^{(1)}&=\frac{32L^6\cos^2\theta\sin\theta(11-\cos2\theta)}{27(3-\cos2\theta)}\cosh\rho\sinh^3\rho\sin^2\phi_1\sin\phi_2,\nonumber\\
F_7^{(2)}&=\frac{4}{9}L^6(\cos3\theta-5\cos\theta)\cosh\rho\sinh^3\rho\sin^2\phi_1\sin\phi_2,\nonumber\\
F_7^{(3)}&=-\frac{256L^6\cos^2\theta\sin\theta}{27(3-\cos2\theta)}\cosh\rho\sinh^3\rho\sin^2\phi_1\sin\phi_2.
\end{align}
Finally, $F_1=F_9=0$ are the field strengths corresponding to the dilaton and the axion fields.

\section{Dual R-R field strengths}\label{sec:AppB}

In this appendix we store some intermediate formulas and calculations, which are drawn from the main text because they are too technical or do not contribute to the main idea.

The expansion of the exponent in formula \eqref{eq:f-m} in powers of the gauge invariant flux $\mathcal{F}$ (up to fourth order) produces terms, which are first order in the gauge invariant flux,
\begin{equation}\label{eq:F1}
\mathcal{F}=A^{(2,0)}+A^{(1,1)}+A^{(0,2)}+A^{(1,0)}+ A^{(0,0)}+A^{(0,1)},
\end{equation}
terms, which are second order in the gauge invariant flux,
\begin{align}\label{eq:F2}
\mathcal{F}^2&=\frac{1}{2}A^{(1,1)}\wedge A^{(1,1)}+\frac{1}{2}A^{(0,1)}\wedge A^{(0,1)}+A^{(1,1)} \wedge A^{(2,0)}+A^{(1,0)}\wedge A^{(2,0)}+A^{(0,2)}\wedge A^{(2,0)}\nonumber\\
&+A^{(1,0)}\wedge A^{(1,1)}+A^{(0,1)}\wedge A^{(2,0)}+A^{(0,0)}\wedge A^{(2,0)}+A^{(0,2)}\wedge A^{(1,1)}+ A^{(0,2)}\wedge A^{(1,0)}\nonumber\\
&+A^{(0,1)}\wedge A^{(1,1)}+A^{(0,1)}\wedge A^{(1,0)}+A^{(0,0)}\wedge A^{(1,1)}+A^{(0,1)}\wedge A^{(0,2)}+ A^{(0,0)}\wedge A^{(0,2)},
\end{align}
terms, which are third order in the gauge invariant flux,
\begin{align}\label{eq:F3}
\mathcal{F}^3&=\frac{1}{6}A^{(1,1)}\wedge A^{(1,1)}\wedge A^{(1,1)}+\frac{1}{2}A^{(1,0)}\wedge A^{(1,1)}\wedge A^{(1,1)}+\frac{1}{2}A^{(0,1)}\wedge A^{(1,1)}\wedge A^{(1,1)}\nonumber\\
&+\frac{1}{2}A^{(0,0)}\wedge A^{(1,1)}\wedge A^{(1,1)}+\frac{1}{2}A^{(0,1)}\wedge A^{(0,1)}\wedge A^{(2,0)}+\frac{1}{2}A^{(0,1)}\wedge A^{(0,1)}\wedge A^{(1,1)}\nonumber\\
&+A^{(0,2)}\wedge A^{(1,1)}\wedge A^{(2,0)}+A^{(0,2)}\wedge A^{(1,0)}\wedge A^{(2,0)}+A^{(0,1)}\wedge A^{(1,1)}\wedge A^{(2,0)}\nonumber\\
&+A^{(0,1)}\wedge A^{(1,0)}\wedge A^{(2,0)}+A^{(0,0)}\wedge A^{(1,1)}\wedge A^{(2,0)}+ A^{(1,0)}\wedge A^{(0,2)}\wedge A^{(1,1)}\nonumber\\
&+A^{(0,1)}\wedge A^{(0,2)}\wedge A^{(2,0)}+A^{(0,1)}\wedge A^{(1,0)}\wedge A^{(1,1)}+A^{(0,0)}\wedge A^{(0,2)}\wedge A^{(2,0)}\nonumber\\
&+A^{(0,1)}\wedge A^{(1,0)}\wedge A^{(0,2)}+A^{(0,0)}\wedge A^{(0,2)}\wedge A^{(1,1)},
\end{align}
and terms, which are fourth order in the gauge invariant flux,
\begin{align}\label{eq:F4}
\mathcal{F}^4&=\frac{1}{6}A^{(0,0)}\wedge A^{(1,1)}\wedge A^{(1,1)}\wedge A^{(1,1)}+\frac{1}{2}A^{(0,1)}\wedge A^{(1,0)}\wedge A^{(1,1)}\wedge A^{(1,1)}\nonumber\\
&+\frac{1}{2}A^{(0,1)}\wedge A^{(0,1)}\wedge A^{(1,1)}\wedge A^{(2,0)}+A^{(0,1)}\wedge A^{(1,0)}\wedge A^{(0,2)}\wedge A^{(2,0)}\nonumber\allowdisplaybreaks[0]\\
&+A^{(0,0)}\wedge A^{(0,2)}\wedge A^{(1,1)}\wedge A^{(2,0)}.
\end{align}

The next step is to gather all differential forms that contain wedge products of three, two, one, and zero $\sigma_a$'s, wedge them by $\mathcal{G}^{(0)}$, $\mathcal{G}_a^{(1)}$, $\mathcal{G}_{ab}^{(2)}$, and $\mathcal{G}^{(3)}$ respectively, and integrate the product along the fiber. The results are listed below (many terms vanish due to repeating differentials).
\begin{itemize}
\item Terms proportional to $\mathcal{G}^{(0)}$:
\begin{align}\label{eq:G0}
F_5^{(1)}\text{vol}(\text{AdS}_5)\wedge\left\{\left[B_{\theta\sigma_1}\widehat{B}_{\phi x_1}+\widehat{B}_{\theta\phi}+ \epsilon_{abc}B_{\theta\sigma_1}\widehat{B}_{\phi x_a}\widehat{B}_{x_bx_c}\left(2x_1+B_{\sigma_2\sigma_3}\right)\right.\right.\nonumber\allowdisplaybreaks[0]\\
\left.\left.\!+2B_{\theta\sigma_1}B_{\phi\sigma_2}\widehat{B}_{x_1x_2}+2\widehat{B}_{\theta\phi}\left(\epsilon_{abc} \widehat{B}_{x_ax_b}x_c+\widehat{B}_{x_2x_3}B_{\sigma_2\sigma_3}\right)\right]\dd\theta\wedge\dd\phi\wedge\dd x_1\wedge\dd x_2\wedge\dd x_3\right.\nonumber\allowdisplaybreaks[0]\\
\left.+\left[1+2\left(\epsilon_{abc}\widehat{B}_{x_ax_b}x_c+\widehat{B}_{x_2x_3}B_{\sigma_2\sigma_3}\right)\right]\dd x_1\wedge\dd x_2\wedge\dd x_3\right.\nonumber\\
\left.+\widehat{B}_{\phi x_a}\left(2x_b\dd x_b+B_{\sigma_2\sigma_3}\dd x_1\right)\wedge\dd\phi\wedge\dd x_a+ \left(B_{\theta\sigma_1}\dd\theta\wedge\dd x_2-B_{\phi\sigma_2}\dd\phi\wedge\dd x_1\right)\wedge\dd x_3\right.\nonumber\\
\left.+\widehat{B}_{x_ax_b}\left(2B_{\theta\sigma_1}x_1\dd\theta+B_{\theta\sigma_1}B_{\sigma_2\sigma_3}\dd\theta+ 2B_{\phi\sigma_2}x_2\dd\phi\right)\wedge\dd x_a\wedge\dd x_b\right.\nonumber\\
\left.+B_{\theta\sigma_1}\widehat{B}_{\phi x_a}\left(2x_1+B_{\sigma_2\sigma_3}\right)\dd\theta\wedge\dd\phi\wedge\dd x_a+ B_{\theta\sigma_1}B_{\phi\sigma_2}\dd\theta\wedge\dd\phi\wedge\dd x_3\right.\nonumber\\
\left.+\left(2x_a\dd x_a+B_{\sigma_2\sigma_3}\dd x_1\right)+\widehat{B}_{\theta\phi}\dd\theta\wedge\dd\phi \wedge\left(2x_a\dd x_a+B_{\sigma_2\sigma_3}\dd x_1\right)\right.\nonumber\\
\left.+\left(2B_{\theta\sigma_1}x_1\dd\theta+B_{\theta\sigma_1}B_{\sigma_2\sigma_3}\dd\theta+2B_{\phi\sigma_2}x_2 \dd\phi\right)\vphantom{\widehat{B}_{\phi x_1}}\right\}.
\end{align}

\item Terms proportional to $\mathcal{G}_a^{(1)}$:
\begin{align}\label{eq:G1}
\left\{\left[-\widehat{B}_{\phi x_1}+\left(x_1-B_{\sigma_2\sigma_3}\right)\epsilon_{abc}\widehat{B}_{\phi x_a} \widehat{B}_{x_bx_c}\right]F_7^{(2)}-\left(B_{\phi\sigma_2}F_7^{(2)}+B_{\theta\sigma_1}F_7^{(3)}\right)\epsilon_{3ab} \widehat{B}_{x_ax_b}\right\}\nonumber\\
\times\text{vol}(\text{AdS}_5)\wedge\dd\theta\wedge\dd\phi\wedge\dd x_1\wedge\dd x_2\wedge\dd x_3\nonumber\\
-\left[\frac{1}{2}\epsilon_{1ab}+\left(2x_1+B_{\sigma_2\sigma_3}\right)\widehat{B}_{x_ax_b}\right]\left(F_3^{(3)}\dd\theta \wedge\dd\phi+ F_7^{(2)}\text{vol}(\text{AdS}_5)\wedge\dd\theta\right)\wedge\dd x_a\wedge\dd x_b\nonumber\\
-\left[\frac{1}{2}\epsilon_{2ab}+2x_2\widehat{B}_{x_ax_b}\right]F_7^{(3)}\text{vol}(\text{AdS}_5)\wedge\dd\phi\wedge\dd x_a\wedge\dd x_b\nonumber\\
-\left(2x_1+B_{\sigma_2\sigma_3}\right)\widehat{B}_{\phi x_a}F_7^{(2)}\text{vol}(\text{AdS}_5)\wedge\dd\theta\wedge\dd\phi \wedge\dd x_a\nonumber\\
-\left(B_{\phi\sigma_2}F_7^{(2)}+B_{\theta\sigma_1}F_7^{(3)}\right)\text{vol}(\text{AdS}_5)\wedge\dd\theta\wedge\dd\phi \wedge\dd x_3\nonumber\\
-\left(2x_1+B_{\sigma_2\sigma_3}\right)\left(F_3^{(3)}\dd\theta\wedge\dd\phi+F_7^{(2)}\text{vol}(\text{AdS}_5)\wedge\dd\theta\right)-2x_2F_7^{(3)}\text{vol}(\text{AdS}_5)\wedge\dd\phi.
\end{align}

\item Terms proportional to $\mathcal{G}_{ab}^{(2)}$:
\begin{align}\label{eq:G2}
-\left[\epsilon_{abc}B_{\theta\sigma_1}\widehat{B}_{\phi x_a}\widehat{B}_{x_bx_c}+\epsilon_{1ab}\widehat{B}_{\theta\phi} \widehat{B}_{x_ax_b}\right]F_7^{(1)}\text{vol}(\text{AdS}_5)\wedge\dd\theta\wedge\dd\phi\wedge\dd x_1\wedge\dd x_2\wedge\dd x_3\nonumber\\
+\widehat{B}_{x_ax_b}\left[\epsilon_{2ab}F_3^{(1)}\dd\theta-\epsilon_{1ab}\left(F_3^{(2)}\dd\phi+F_7^{(1)}\text{vol}(\text{AdS}_5)\right)\right]\wedge\dd x_1\wedge\dd x_2\wedge\dd x_3\nonumber\\
+\widehat{B}_{\phi x_a}F_3^{(1)}\dd\theta\wedge\dd\phi\wedge\dd x_a\wedge\dd x_2-\widehat{B}_{\phi x_a}F_7^{(1)}\text{vol}(\text{AdS}_5)\wedge\dd\phi\wedge\dd x_a\wedge\dd x_1\nonumber\\
-\widehat{B}_{x_ax_b}\left[\left(B_{\phi\sigma_2}F_3^{(1)}+B_{\theta\sigma_1}F_3^{(2)}\right)\dd\phi+ B_{\theta\sigma_1}F_7^{(1)}\text{vol}(\text{AdS}_5)\right]\wedge\dd\theta\wedge\dd x_a\wedge\dd x_b\nonumber\\
-B_{\theta\sigma_1}\widehat{B}_{\phi x_a}F_7^{(1)}\text{vol}(\text{AdS}_5)\wedge\dd\theta\wedge\dd\phi\wedge\dd x_a- \widehat{B}_{\theta\phi}F_7^{(1)}\text{vol}(\text{AdS}_5)\wedge\dd\theta\wedge\dd\phi\wedge\dd x_1\nonumber\\
+F_3^{(1)}\dd\theta\wedge\dd x_2-\left(F_3^{(2)}\dd\phi+F_7^{(1)}\text{vol}(\text{AdS}_5)\right)\wedge\dd x_1\nonumber\\
-\left[\left(B_{\phi\sigma_2}F_3^{(1)}+B_{\theta\sigma_1}F_3^{(2)}\right)\dd\phi+B_{\theta\sigma_1}F_7^{(1)}\text{vol}(\text{AdS}_5)\right]\wedge\dd\theta.
\end{align}

\item Terms proportional to $\mathcal{G}^{(3)}$:
\begin{equation}\label{eq:G3}
F_5^{(2)}\widehat{B}_{x_ax_b}\dd\theta\wedge\dd\phi\wedge\dd x_a\wedge\dd x_b+F_5^{(2)}\dd\theta\wedge\dd\phi.
\end{equation}
\end{itemize}
In the above formulas the sum over all repeating indices is implied for $a,b,c=1,2,3$.

\end{appendix}

\end{document}